\documentclass[twocolumn,aps,prc]{revtex4}
\usepackage{amssymb}
\usepackage{amsmath}
\usepackage{color}
\usepackage{graphicx,multirow,dcolumn}
\usepackage{caption}
\usepackage[english]{babel}
\usepackage{epstopdf}
\def\babar{\mbox{\slshape B\kern-0.1em{\smaller A}\kern-0.1em
    B\kern-0.1em{\smaller A\kern-0.2em R}}}
\newcommand{\er}{$\pm$}
\newcommand{\beq}{\begin{eqnarray}}
\newcommand{\eeq}{\end{eqnarray}}
\newcommand{\bc}{\begin{center}}
\newcommand{\ec}{\end{center}}
\newcommand{\bi}{\begin{enumerate}}
\newcommand{\ei}{\end{enumerate}}
\newcommand{\bit}{\begin{itemize}}
\newcommand{\eit}{\end{itemize}}
\newcommand{\rd}{\color{darkred}}
\newcommand{\bl}{\color{darkblue}}

\newcommand{\gr}{\color{darkgreen}}

\definecolor{lightyellow}{cmyk}{0,0,0.5,0}
\definecolor{lightred}{rgb}{1,0.5,0.5}
\definecolor{lightgreen}{rgb}{0,0.4,0}
\definecolor{lightblue}{rgb}{0.5,0.5,1}
\definecolor{darkred}{rgb}{0.8,0,0}
\definecolor{darkgreen}{rgb}{0,0.4,0}
\definecolor{darkcyan}{cmyk}{1,0.3,0.3,0.3}
\definecolor{darkblue}{rgb}{0,0,0.6}
\definecolor{lightbrown}{rgb}{0.7,0.3,0.3}
\definecolor{darkbrown}{rgb}{0.5,0,0}
\begin{document}

\title{Are the XYZ states unconventional states or conventional states with unconventional properties?}

\newcommand*{\HISKP}{Helmholtz--Institut f\"ur Strahlen- und Kernphysik, Universit\"at Bonn, 53115 Bonn, Germany}
\newcommand*{\IKP}{Institute for Andvanced Simulation and Institut f\"ur Kernphysik, Forschungszentrum J\"ulich, J\"ulich, Germany}
\affiliation{\IKP}
\affiliation{\HISKP}

\author{Christoph Hanhart} \affiliation{\IKP}
\author{Eberhard~Klempt} \affiliation{\HISKP}

\begin{abstract}
We discuss three possible scenarios for the interpretation of mesons containing a heavy quark and its
antiquark near and above the first threshold for a decay into a pair of heavy mesons in a relative $S$--wave. 
View I assumes that these thresholds force the quark potential to flatten which implies that while
in these energy ranges molecular states may be formed there should not be any quark--anti-quark states
above these thresholds.
 View II assumes that the main part of the
interaction between two mesons is due to the poles which originate from the $Q\bar Q$ interaction. The 
properties of the $Q\bar Q$ mesons are strongly influenced by opening thresholds but the number of states
is given by the quark model. In View III, both types of mesons are admitted also near and above 
the open flavor thresholds: $Q\bar Q$ mesons and dynamically
generated mesons. Experimental consequences of these different views are discussed.  
\end{abstract}

\date{\today}


\maketitle

\section{Introduction}
Great progress has been achieved in the spectroscopy of hadrons containing two heavy
quarks due to the tremendous efforts of experiments like BaBar, Belle, BESIII, CLEO, LHCb,
$\cdots$, and further progress is expected from the ongoing programs and, in the future,
from Belle II and PANDA.
At present, the Particle Data Group (PDG)~\cite{Tanabashi:2018oca} lists 37 states containing a
$c\bar c$ and 20 states containing a $b\bar b$ pair.  Amongst those there are
many states with unexpected properties, like~\cite{footnote1} $\chi_{c1}(3872)$ also known 
as (aka) $X(3872)$,
$\psi(4260)$ aka $Y(4260)$, $\psi(4360)$ aka $Y(4360)$, $\psi(4660)$ aka $Y(4660)$, and
$\chi_{c1}(4140)$ and $\chi_{c1}(4274)$. Moreover, there are even states with isospin $I=1$ (established
are $Z_c(3900)$, $Z_c(4020)$, $Z_c(4430)$, $Z_b(10610)$, $Z_b(10650)$) decaying to final
states that contain a heavy quark and its antiquark --- as such the states must contain
at least four quarks.
All these states are classified as
{\it unconventional} states or as {\it candidates for an exotic structure}, but it is unclear
what their underlying structure is.

In the literature those states are typically proposed to be 
quarkonia ($Q\bar Q$), possibly with unconventional properties, compact tetraquarks 
(diquark---antidiquark $(qQ)$-$(\bar Q\bar q)$), hybrids ($\bar QQ$ states with active gluons
contributing to the quantum numbers),  hadroquarkonia (with a
structure as  ($Q\bar Q$)-($q\bar q$)), or loosely bound molecular states ($Q\bar
q$)-($Q\bar q$). A large number of reviews has appeared recently that discuss the exotic
candidates from different angles, see, e.g.,
Refs.~\cite{Brambilla:2010cs,Esposito:2016noz,Chen:2016qju,Ali:2017jda,Lebed:2016hpi,Olsen:2017bmm,Guo:2017jvc}.
The key issue is if in the presence of light quarks the heavy quark--antiquark potential keeps rising as it does in the quenched approximation
of the potential. This would imply that near and above the first relevant $S$--wave open flavor threshold
at most molecular states could exist but no quark--anti-quark states.
 It should be stressed that many unexpected phenomena were discovered very
close to important thresholds.

The problem at hand is probably best explained by a brief look at $\chi_{c1}(3872)$.
Its very small binding energy (there is currently only an upper limit of 180 keV for
this binding energy) makes this state a prime example of a
loosely bound molecule. However, the question remains, if this state is just a molecule produced by
two-hadron interactions or if it owes it existence a $c\bar c$ core.
The $\chi_{c1}(2P)$ may still be waiting for discovery --- or it is already found and
should be identified with the $\chi_{c1}(4140)$. The pattern of the
$\chi_c(1P)$ states suggests that the three states $\chi_{c2}(3930)$, $\chi_{c1}(3872)$, $\chi_{c0}(3860)$
could be the $\chi_c(2P)$  states. In this paper, we compare the
implications of three very different hypotheses regarding the doubly heavy states near or above the
first relevant open heavy flavor threshold.

{\bf View I} underlines the importance of the
``molecular" interaction between two mesons. In this view, the $\chi_{c1}(3872)$ is an isoscalar
$D^*\bar D + c.c.$ molecule unrelated to the $c\bar c$ system. (The charge conjugated component is
omitted from now onwards.) Here, as in all partial waves, quarkonia exist only below the first 
relevant $S$-wave threshold for a
two-particle decay --- this statement implies that these two particles must be narrow, $\Gamma\ll
\Lambda_{\rm QCD}$, for otherwise the possible molecule would be too broad~\cite{Filin:2010se} or,
stated differently, would have already decayed before it could hadronize~\cite{Guo:2011dd}. In this
view it is assumed that at this threshold virtual light quarks screen the quark-antiquark
potential. As a result the potential flattens off and all resonances at or above the threshold are
of molecular nature. In this scenario, $Q\bar Q$ states exist only below this threshold, and 
the number of molecular states is (at most) given by the number of
relevant $S$--wave thresholds in the kinematic range of interest (although there might also be
$P$--wave states observed already --- this is discussed  below). Note that not necessarily
all $S$--wave channels have a sufficiently strong attractive interaction to generate singularities
with a significant impact on observables (Note that in the two nucleon sector there is a bound
state only in the spin triplet, isospin singlet channel. In the spin singlet, isospin triplet channel
there is only a virtual state which is, however, so close to the threshold that it generates
a very large scattering length).

{\bf View II} is based on the assumption that the leading part of the interaction between two mesons
is due to their $Q\bar Q$ component.
The argument is that there can be different reasons to expect a resonance in a given mass range.
Mesons with a given set of quantum numbers can be $q\bar q$, they could be hybrids (abbreviated
often as $q\bar q g$), tetraquarks $qq\bar q\bar q$, molecular meson-meson resonances, 
baryonia (baryon-antibaryon bound states or resonances) or glueballs. These are
six different possible species. However, there is no experimental evidence
for such an abundance. View II assumes that
these different ingredients may be components in the mesonic wave function, but that these options do not manifest
themselves in separate resonances. In this view, the number of expected heavy-quark states is given by the
number of expected $Q\bar Q$ states. It is assumed that these states drive the major part of the interactions
between the particles into which the states decay. Due to threshold openings, the properties
of the wave function can change as well as the resonance parameters but not the number of states.
In this view, $\chi_{c1}(3872)$ would have a  $\bar QQ$ {\it and} a sizeable
molecular component. But there is one $\chi_{c1}$  state only in this mass range that should be
identified with $\chi_{c1}(2P)$.

One can in principle also think of a mixture of View I and II, if one were to admit that
deuteron-like loosely bound states of two hadrons might exist if there is no possibility to reduce
the number of quarks. In this formulation, exotic mesons like the $Z$ particles ($Z_c(3900)$,
$Z_b(10610)$, $\cdots$) might exist as poles of the $S$--matrix. However, we will not 
go deeper into this discussion here.

Finally, in {\bf View III}, we allow for the existence of both even above the first relevant
two--hadron threshold: States that owe their existence their $\bar QQ$ core as well as those that are
of molecular nature. In this scenario the number of states will exceed
the number of states defined by the $\bar QQ$ model. Moreover, one
expects states near $S$--wave thresholds as well as states with masses unrelated to those.

In principle the same issues raised above could also be discussed for the light quark sector, however,
due to the non-perturbative nature of QCD at small momentum transfers but asymptotic freedom at
large scales, one expects that heavy--heavy systems, which are the focus of this work, are
easier to analyse than heavy--light or all--light systems.
Moreover, the heavy quark spin symmetry (HQSS) states that, up to corrections of order
$\Lambda_{\rm QCD}/M_Q$ where $\Lambda_{\rm QCD}\simeq 200$ MeV denotes the QCD mass scale
 and $M_Q$ the heavy quark mass, the heavy quark spin 
does not interact. This results in the appearance of spin multiplets and 
allows one to identify selection rules for certain decays that are
sensitive to the internal structure of the states, both of which proved to be  important diagnostic tools
 when it comes to classifying 
exotic states.
In addition, mesons have an easier
substructure than baryons. Thus in what follows we focus on doubly heavy mesonic systems. 

\section{The bottomonium spectrum}

\begin{figure}[t]
\includegraphics[width=0.48\textwidth]{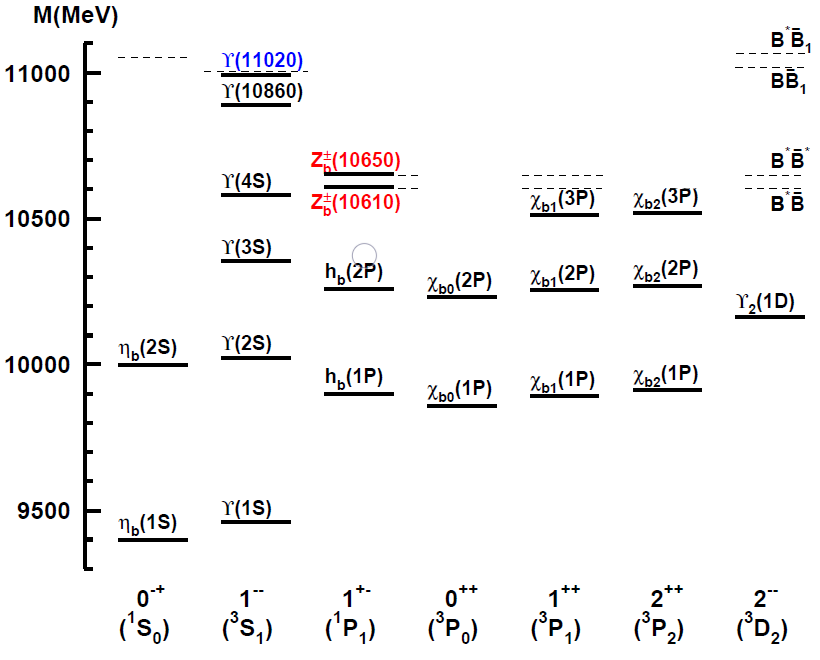}
\caption{\label{bbbar}(Color online) The bottomonium states. Established states are given by a solid line
and their names. The thresholds for $S$-wave decays in each channel
are indicated by dashed lines. The $1^{++}$ and $1^{+-}$ partial waves have their $S$-wave
threshold at the $B^*\bar B$, the $2^{++}$ partial wave at the $B^*\bar B^*$ threshold.   
In View I, states close to or above thresholds (in blue) and
the isospin 1 states (in red) are interpreted as molecular states. In View II only states compatible
with $Q\bar Q$ exist.}
\end{figure}

Fig.~\ref{bbbar}
shows the spectrum in the b-quark sector. The spectrum is very clean. There is a series of
$\Upsilon$ states, $\Upsilon(1S)\cdots \Upsilon(4S)$~\cite{footnote2}, $\Upsilon(10860)$ and $\Upsilon(11020)$, with quantum
numbers $I^G(J^{PC})=0^+(1^{--}$) where $I$, $G$, $J$, $P$, $C$ are the isospin, G-parity, total
spin, parity and C-parity of the mesons. The vector states can be produced in $e^+e^-$
annihilation, and most of our detailed knowledge on the $\Upsilon$-family of states stems from this
process. The $\Upsilon(n^3D_1)$ states have the same quantum numbers as the $\Upsilon(nS)$ states
and could in principle be produced in $e^+e^-$ annihilation as well, but this production violates
spin symmetry which is most probably the reason why those have not been seen here. The
$\Upsilon(1^3D_2)$ state (with orbital angular momentum $L=2$ and quark spin $S=1$) has been 
seen in a $\Upsilon(3S)\to
\gamma \chi_{b}(2P)$, $\chi_{b}(2P)\to \gamma \Upsilon(1^3D_2)$, $\Upsilon(1^3D_2)\to \gamma
\chi_{b}(1P)$, $\chi_{b}(1P)\to \gamma \Upsilon(1S)$ cascade decay with four photons in the final
state~\cite{Bonvicini:2004yj}.

The two resonances $\Upsilon(10860)$ and $\Upsilon(11020)$ are above the open beauty threshold ---
given the quantum numbers, the decay $\Upsilon\to B^{(*)}\bar B^{(*)}$ happens in a $P$-wave. The
mass of the  $\Upsilon(11020)$ is right below the first $S$-wave threshold, namely $B_1\bar B$,
where $B_1$ denotes the axial vector $B$-meson with the light quark cloud carrying $j=3/2$.

Further states are known: There are two pseudoscalar mesons, $\eta_b(1S)$ and  $\eta_b(2S)$. They
are found slightly below the corresponding vector states in line with expectations from
HQSS for $\bar bb$ states. In addition, there are two complete
quartets with $L=1$; two spin triplets $\chi_{bJ}(nP)$ with $J=0,1,2$ and $n=1,2$ and two spin
singlets $h_b(nP)$, again with $n=1,2$. Two  states belong to the $3P$ series: $\chi_{b1}(3P)$
and the recently discovered $\chi_{b2}(3P)$~\cite{Sirunyan:2018dff}.

The spin-triplet and spin-singlet  states satisfy the center-of-gravity rule which holds true
 when tensor and spin-spin forces are negligible:
\begin{eqnarray}
\hspace{-2mm}M_{h_{Q}(nP)} =\frac19 \left(5 M_{\chi_{Q2}(nP)} + 3 M_{\chi_{Q1}(nP)} + M_{\chi_{Q0}(nP)}\right)
\label{cog}
\end{eqnarray}
For $Q$\,=\,$b,n$\,=\,1, the difference between the left hand side and the right hand side is $\delta M$\,=\,-(0.57$\pm$1.08)\,MeV and for $Q$\,=\,$b,n$\,=\,2, 
$\delta M$\,=\,$-(0.4\pm1.3)$\,MeV.  
The center-of-gravity rule is excellently satisfied.

Note that with the exception of $\Upsilon(11020)$ all states discussed so far are well  below the
threshold for $S$-wave decays. The pertinent thresholds for the different quantum numbers for
$S$-wave decays are shown in Table~\ref{tab:Swavethresholds}.

\begin{table}
\caption{The lowest $S$--wave thresholds for a given $J^{PC}$, shown in the first
line.\label{tab:Swavethresholds}}
\renewcommand{\arraystretch}{1.4}
\bc
\begin{tabular}{ccccccc}
\hline\hline
 $0^{-+}$    & $1^{--}$  & $1^{+-}$  & $0^{++}$& $1^{++}$  & $2^{++}$     & $2^{--}$\\
\hline
         $B_1\bar B^*$&$B_1\bar B$&$B^*\bar B$&$B\bar B$&$B^*\bar B$&$B^*\bar B^*$&$B_2\bar B$\\
          11.050      & 11.004    & 10.604    & 10.558  & 10.604    & 10.650 & 11.019\\
         \hline\hline
        \end{tabular}\ec
\renewcommand{\arraystretch}{1.}
 \end{table}

The two isotriplets of states $Z_b(10610)$ and $Z_b(10650)$ with quantum numbers $1^+(1^{-+})$ are
evidently not $b\bar b$ mesons and have no pure  $b\bar b$  component. The minimal quark content
for a $Z_b^+$ is $b\bar bu\bar d$ with four quarks suggesting a tetraquark
configuration~\cite{Ali:2011ug}.  However, $Z_b(10610)$ and $Z_b(10650)$ decay not only into
bottomonium states, they also decay into pairs of mesons with open bottomness: $Z_b(10610)$ with a
fraction of $(82.6\pm2.9\pm2.3)$\% into $\bar BB^*$, and $Z_b(10650)$ with $(70.6\pm4.9\pm4.4)$\%
into $\bar B^*B^*$ (but not into $\bar BB^*$). Thus a molecular nature of $Z_b(10610)$ and
$Z_b(10650)$ is very likely~\cite{Bondar:2011ev} even though a kinematical origin~\cite{Bugg:2011jr,Swanson:2014tra} is not yet fully excluded.
 Both mesons are very close to a threshold. The
PDG quotes~\cite{Tanabashi:2018oca}
\begin{eqnarray} \nonumber
M_{Z_b(10610)^\pm}-(M(B^*) + M(B))\phantom{^*}&=&4{\pm}2  \ \mbox{MeV} \\
M_{Z_b(10650)^\pm}-(M(B^*) + M(B^*))&=&4{\pm}1.5 \ \mbox{MeV} \nonumber
\end{eqnarray}
A study of their line shape~\cite{Wang:2018jlv} finds that the poles related to the two $Z_b$ are
located even closer to the corresponding thresholds, but on the unphysical Riemann sheets. In
particular, the $Z_b(10610)$ is found as virtual state (just below the $B^*\bar B$ threshold) while
the $Z_b(10650)$ pole is found just above the $B^*\bar B^*$ threshold. It is not difficult to
anticipate that the next charged pair of $Z_b$ states can be expected at the $\bar B_sB_s^*$ and
$\bar B_s^*B_s^*$ thresholds, at 10782 and 10831\,MeV. They could be produced in $\Upsilon(11020)$
decays. All these isovector states are evidently not of $q\bar q$ nature. Their observation is
contrasted with the different views below.

\section{Charmonium\vspace{-2mm}}

\begin{figure}[pt]
\hspace{-6mm}
\includegraphics[width=0.50\textwidth,height=0.383\textwidth]{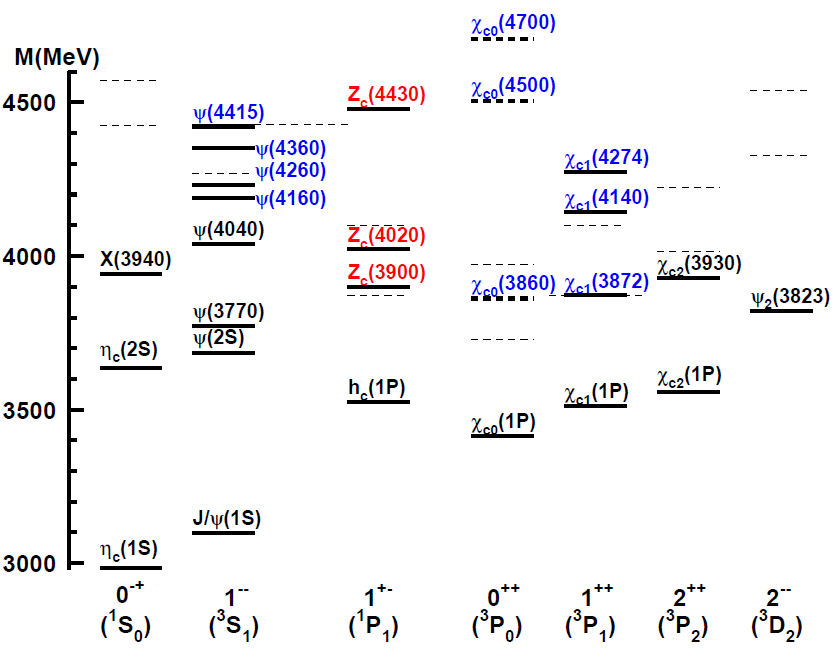}
\caption{\label{ccbar}(Color online) The charmonium states. See also caption
of Fig.~\ref{bbbar}. Candidate states are shown by a dashed line. Non-established states with isospin 1
are not shown, see Table~\ref{XZs} for a full list. The $X(4020)$ is shown as $Z_c(4020)$. 
The thresholds for $S$-wave decays into a $c\bar n-n\bar c$ and
$c\bar s-s\bar c$ in each partial wave are depicted as thin dashed lines.  
 $X(3940)$ is assumed to have pseudoscalar quantum numbers and is interpreted as
$\eta_c(3S)$.  \vspace{-2mm}}
\end{figure}

\begin{table}[pt]
\caption{\label{ccb-low} The spectrum of charmonium states below the $D\bar D$
threshold.\vspace{-2mm} } \setlength{\tabcolsep}{0.23pc}
\begin{center}
\begin{tabular}{lccccccc}
\hline\hline
\rule[10pt]{-1mm}{0mm}
 State & $m$~(MeV) & $\Gamma$~(MeV) & $I^G(J^{PC})$ \\[0.7mm]
\hline
\rule[10pt]{-1mm}{0mm}
$\psi(1S)$      &$3096.900\pm0.006$&$0.0929\pm0.0028$& $0^+(1^{--})$\\[1mm]
$\chi_{c0}(1P)$ & $3414.71\pm0.30$ & $10.8\pm0.6$    & $0^+(0^{++})$\\[1mm]
$\chi_{c1}(1P)$ & $3510.67\pm0.05$ & $0.84\pm0.04$   & $0^+(1^{++})$\\[1mm]
$h_c(1P)$       & $3525.45\pm0.15$ & $0.70\pm0.40$   & $0^+(1^{+-})$\\[1mm]
$\chi_{c2}(1P)$ & $3556.17\pm0.07$ & $1.97\pm0.09$   & $0^+(2^{++})$\\[1mm]
$\eta_c(2S)$    & $3637\pm4$       &14$\pm$7         & $0^+(0^{-+})$\\[1mm]
$\psi(2S)$      &$3686.097\pm0.025$&$0.294\pm8$      & $0^+(1^{--})$\\
\hline\hline
\end{tabular}
\end{center}
\vspace{-5mm}
\end{table}
Figure~\ref{ccbar} shows the charmonium states listed by the PDG~\cite{Tanabashi:2018oca} which
contain a $c\bar c$ pair in their wave function. States up to $\psi(2S)$ have masses below the open
charm ($D\bar D$) threshold and are narrow, mostly with a width of a few MeV or even smaller.
All expected charmonium states below the $D\bar
D$ threshold are known and unambiguously established. They are collected in Table~\ref{ccb-low}.
The center-of-gravity rule, Eq.~(\ref{cog}), holds true for $Q=c,n=1$ with $\delta M= 0.08\pm 0.61$\,MeV. 
\begin{table}[pb]
\caption{\label{thresholds}Thresholds for production two charmed mesons. The quantum numbers
correspond to $S$-wave decays. }
\bc\small
\renewcommand{\arraystretch}{1.2}
\begin{tabular}{lccccccc}\hline\hline\\[-3ex]
$J^{PC}$ & $0^{-+}$    & $1^{--}$  & $1^{+-}$  & $0^{++}$& $1^{++}$  & $2^{++}$     & $2^{--}$\\\hline
         &$D_1\bar D^*$&$D_1\bar D$&$D^*\bar D$&$D\bar D$&$D^*\bar D$&$D^*\bar D^*$&$D_2^*\bar D$\\
         & 4.4228      & 4.286   & 3.872    & 3.730  & 3.872   & 4.014 & 4.326\\\hline
         &$D_{s1}\bar D_s ^*$&$D_{s1}\bar D_s$&$D_s ^*\bar D_s$&$D_s\bar D_s$&$D_s ^*\bar D_s$&$D_s ^*\bar D_s ^*$&$D_{2s}\bar D_s$\\
         & 4.572      & 4.428    & 4.081    & 3.972  & 4.081    & 4.224 & 4.537\\\hline\hline
\end{tabular}
\renewcommand{\arraystretch}{1.}
\ec
\end{table}

The states above the open charm threshold are significantly broader. Here, two thresholds become
important: The threshold for $S$-wave decays into a $c\bar n$-$n\bar c$ pair --
where we include couplings to the ground state $D$ mesons and the narrow even-parity $D$-mesons -- and into  
$c\bar s$-$s\bar c$ are shown in Table~\ref{thresholds}. The full charmonium
spectrum is displayed in Fig.~\ref{ccbar}, where the thresholds are shown as dashed lines.

The PDG lists ten $\psi$ states but Fig.~\ref{ccbar} shows only eight: There is the
well known $\psi(4260)$, seen in the $J/\psi\pi\pi$~\cite{Ablikim:2016qzw}, $J/\psi K\bar
K$~\cite{Ablikim:2018epj}and $\pi^+D^{0}D^{*-}$ \cite{Ablikim:2018vxx} final states, 
and the candidate state $\psi(4230)$ observed to decay into
 $\pi\pi h_c$~\cite{BESIII:2016adj}, $\omega\chi_{co}$ \cite{Ablikim:2014qwy}, and
$\pi\pi\psi(2S)$~\cite{Ablikim:2017oaf}. We assume here that these phenomena are related and
correspond to one particle in line with the analysis of Ref.~\cite{Cleven:2013mka,Cleven:2016qbn};
this finds further support in the fact that the most recent data for $e^+e^-\to J\psi
\pi^+\pi^-$~\cite{Ablikim:2016qzw} clearly peak between $4220$ and $4230$. Likewise, we identify
$\psi(4390)$, seen in $\pi\pi h_c$~\cite{BESIII:2016adj}, and $\psi(4360)$ decaying into
$\psi(2S)\pi\pi$~\cite{Ablikim:2017oaf}. The four resonance claims, combined here to two states,
are collected in Table~\ref{satellites}.

\begin{table}[pt]
\caption{\label{satellites}BESIII masses and widths (in MeV)
of $\psi$ resonances in the mass range from $\psi(4160)$ to $\psi(4415)$.
$\psi(4260)$ and $\psi(4360)$ are considered to be {\it established} by 
the PDG, $\psi(4230)$ and $\psi(4390)$ not.
}
\renewcommand{\arraystretch}{1.4}
\begin{tabular}{lllll}
\hline\hline
PDG&\qquad Mass &\qquad Width &\quad Decay & Ref.\\
\hline
\multirow{3}{*}{$\psi(4230)$}
&4218.4$_{-4.5}^{+5.5}$\er0.9&\ 66.0$_{-8.3}^{+12.3}$\er 0.4&$\pi^+\pi^-
h_c(1P)$&\cite{BESIII:2016adj}\\
&4230\er 8\er 6              &\ 38\er 12\er 2              &$\omega\chi_{c0}(1P)$&\cite{Ablikim:2014qwy}\\
&4209.5\er 7.4\er 1.4        &\ 80.1\er 24.6\er 2.9         &$\pi^+\pi^-\psi(2S)$&\cite{Ablikim:2017oaf}\\\hline
\multirow{2}{*}{$\psi(4260)$}&4222.0\er 3.1\er 1.4        &\ 44.1\er 4.3\er 2.0          &$\pi^+\pi^- J/\psi$&\cite{Ablikim:2016qzw}\\
&4228.6\er 4.1\er 5.9        &\ 77.1\er 6.8\er 6.9          &$\pi^+D^{0}D^{*-}$ &\cite{Ablikim:2018vxx}\\
\hline
\multirow{1}{*}{$\psi(4360)$}&4320.0\er 10.4\er 7.0      &101.4$_{-19.7}^{+25.3}$\er10.2&$\pi^+\pi^- J/\psi$&\cite{Ablikim:2016qzw}\\
\hline
\multirow{2}{*}{$\psi(4390)$}
&4391.5$_{-6.8}^{+6.3}$\er1.0
&139.5$_{-20.6}^{+16.2}$\er0.6&$\pi^+\pi^-h_c(1P)$&\cite{BESIII:2016adj}\\&4383.8\er 4.2\er 0.8      &\ 84.2\er 12.5\er 2.1        &$\pi^+\pi^-\psi(2S)$&\cite{Ablikim:2017oaf}\\
\hline\hline
\end{tabular}
\end{table}

The Belle collaboration reported a few charmonium states observed in a process in which two $c\bar
c$ pairs are produced in two-photon collisions~\cite{Abe:2007jna}. Three states are
identified with known states:  $\eta_c(1S)$, a weaker $\chi_{c0}(1P)$ decaying into $D\bar D$, and
$\eta_c(2S)$. Two further states are seen, $X(3940)$ decaying to $D^*\bar D$, and $X(4160)$
decaying into $D^*\bar D^*$. Tentatively, we assign the $X(3940)$ state to $\eta_c(3S)$.
If this state were indeed a $0^{-+}$ state, this assignment would in fact be consistent with all
views discussed in this paper, since the lowest lying $S$--wave threshold with these quantum
numbers is at 4.423 MeV ($cf$. Table~\ref{thresholds}).

The Belle collaboration reported the observation of a scalar charmonium state in the reaction
$e^+e^-\to J/\psi D\bar D$ \cite{Chilikin:2017evr}. Its mass was determined to
$(3862^{+26}_{-32}$$^{+40}_{-13})$\,MeV and its width to $(201^{+154}_{-67}$$^{+88}_{-82})$\,MeV.
It is listed as $\chi_{c0}(3860)$ but not included in the PDG summary.

\begin{table}[pt]
\caption{\label{chi2}Measurements on $X(3915)$ and $\chi_{c2}(3930)$ }
\renewcommand{\arraystretch}{1.2}
\begin{center}
\begin{tabular}{lcccc}
\hline\hline
         & Reaction                 &Mass           &Width
                 &  $\Gamma_{f}\cdot\Gamma_{i}/\Gamma_{\rm tot}$ \\
$X(3915)$&$\gamma\gamma\to \omega J/\psi$ &$3918.4\pm 1.9$  & $20\pm 5$   &$\ 54\pm\ 9$\,eV\\
$\chi_{c2}(3930)$&$\gamma\gamma\to D\bar D$       &$3927.2\pm2.6$        & $24\pm6$    &$210\pm40$\,eV \\
\hline\hline
\end{tabular}
\end{center}
\renewcommand{\arraystretch}{1.0}
\end{table}

\begin{table*}[pt]
\caption{\label{XZs}Charmonium states with isospin $I=1$ and the discovery reaction and decay. Close-by thresholds
are given with the corresponding partial wave. Yes/no indicates if a
state is considered to be {\it established} or not by the PDG.}
\renewcommand{\arraystretch}{1.3}
\bc
\begin{tabular}{cccccccccc}
\hline\hline
              & $I^G(J^{PC})$&Mass (MeV)&Width (MeV)&Production&Main decay&Threshold&Wave&Establ. &Ref.\\  \hline
$Z_{c}(3900)$ & $1^+(1^{+-})$&$3918.4\pm1.9$&$20\pm5$&$e^+e^-$ at 4.26\,GeV&$\pi\,J/\psi$&$\bar DD^*$&S&yes&\cite{Tanabashi:2018oca}\\
$X(4020)^\pm$ & $1^+(?^{?-})$&$4024.1\pm1.9$&$13\pm5$&$e^+e^-$ at 3.9-4.42\,GeV&$\pi^\pm h_{c}$$^{a}$&$\bar D^*D^*$&S&yes&\cite{Tanabashi:2018oca}\\
$X(4050)^\pm$ & $1^-(?^{?+})$&$4051^{+24}_{-40}$&$82^{+50}_{-28}$&$\bar B^0\to K^- X^+$&$\pi^\pm \chi_{c1}(1P)$$^{b}$&$\bar D_sD^*_s$&P&no&\cite{Tanabashi:2018oca}\\
$X(4055)^\pm$ & $1^+(?^{?-})$&$4054\pm3.2$&$45\pm13$&$\psi(4260)\to\pi X$&$\pi^\pm \psi(2S)$&$\bar D_sD^*_s$&S&no&\cite{Tanabashi:2018oca}\\
$X(4100)^-$ & $1^+/1^-$&$4096\pm20^{+18}_{-22}$&$152\pm58^{+60}_{-35}$&$B^0\eta_c\to K^+\pi^-$&$X\to\pi^-\eta_c$&&&$3\sigma$ effect&\cite{Aaij:2018bla}\\
$Z_{c}(4200)$ & $1^+(1^{+-})$&$4196^{+35}_{-32}$&$370^{+100}_{-150}$&$\bar B^0\to K^- X^+$&$\pi\,J/\psi$&$\bar D_s^*D_s^*$&P&no&\cite{Tanabashi:2018oca}\\
$R_{c0}(4240)$& $1^+(0^{--})^c$&$4239^{+50}_{-21}$&$220^{+120}_{-90}$&$\bar B^0\to K^- X^+$&$\pi^\pm \psi(2S)$&$\bar D_s^*D^*_s$&P&no&\cite{Tanabashi:2018oca}\\
$X(4250)^\pm$ & $1^-(?^{?+})$&$4248^{+190}_{-50}$&$177^{+320}_{-70}$&$\bar B^0\to K^- X^+$&$\pi^\pm \chi_{c1}(1P)$$^{b}$& $\bar D_1D^*$&S&no&\cite{Tanabashi:2018oca}\\
$Z_{c}(4430)$ & $1^+(1^{+-})$&$4478^{+15}_{-18}$&$181\pm31$&$\bar B^0\to K^- X^+$&$\pi\,\psi(2S)$& $\bar D_{s1}D_s$&P&yes&\cite{Tanabashi:2018oca}\\
\hline\hline\vspace{-6mm}
\end{tabular}
\ec
\renewcommand{\arraystretch}{1.}
$^{a}$ Seen in all three charge states. \qquad $^{b}$ Seen in \cite{Mizuk:2008me}, not seen in
\cite{Lees:2011ik}\\
$^{c}$ The exotic $0^{--}$ quantum numbers are favored over $1^{+-}$ by one $\sigma$\vspace{-3mm}\\
\end{table*}

The PDG identifies the state located near 3930 MeV as $\chi_{c2}(2P)$ state. It is
observed in two-photon collisions \cite{Lees:2012xs} and in $B$ decays \cite{delAmoSanchez:2010jr} 
in its decay into $\omega J/\psi$. Very
close-by is the $X(3915)$ which was formerly identified with $\chi_{c0}(2P)$ since the analysis
favored $J^{PC}=0^{++}$. Table~\ref{chi2} collects the relevant information on $X(3915)$ and
$\chi_{c2}(3930)$. In Ref.~\cite{Zhou:2015uva} it was shown, however, that $X(3915)$ may also have
$J^{PC}=2^{++}$ quantum numbers when the helicity-2 dominance assumed by BABAR is no longer
imposed. 
Thus, the two states may be one single $\chi_{c2}(2P)$ state with a large molecular
component (although current data seems to be compatible with this assignment
only if there are large violations of spin symmetry~\cite{Baru:2017fgv}). Assuming that there is one state only, we evaluate the ratio of branching fractions
\begin{eqnarray}
\frac{{\cal B}\chi_{c2}(2P)\to \omega J/\psi}{{\cal B}\chi_{c2}(2P)\to D\bar D} &=& 0.26\pm 0.07
\end{eqnarray}
The OZI-rule-violating decay into $\omega J/\psi$ is seen with a large branching ratio. The
threshold for the first $S$-wave decay into open charm, into $D^*\bar D^*$, is with 4014\,MeV
quite far away.

The $\chi_{c1}(3872)$ has unconventional properties. Its mass of $3871.69\pm0.17$\,MeV coincides
exactly at the sum of the $\bar D^0$ and $D^{*0}$ masses ($3871.68\pm0.07$\,MeV) and falls below
the sum of the $D^-$ and $D^{*+}$ masses ($3879.91\pm0.07$\,MeV). Hence it decays into $\bar
D^0D^{*0}$ but not into $D^-D^{*+}$. Its  branching ratio for $\gamma \psi(2S)$ is with $4\%$
significantly larger than its branching ratio for $\gamma J/\psi$ which is well below $1\%$. Its
probably most striking feature is, however, that it decays almost equally often into the isovector
final state $J/\psi\pi^+\pi^-$ and into the isoscalar $J/\psi \omega$ final
state~\cite{delAmoSanchez:2010jr}, with
\begin{eqnarray}
\frac{{\cal B}\chi_{c1}(3872)\to \omega J/\psi}{{\cal B}\chi_{c1}(3872)\to \pi^+\pi^- J/\psi} &=& 0.8\pm 0.3.
\end{eqnarray}
\begin{table}[pb]
\caption{\label{scalar}Masses, widths (in MeV) and quantum numbers of $\phi J/\psi$ resonances from
Ref.~\cite{Aaij:2016iza} and possible spectroscopic interpretations.}
\renewcommand{\arraystretch}{1.3}
\begin{center}
\begin{tabular}{ccccc}
\hline\hline
        &$\chi_{c1}(4140)$&$\chi_{c1}(4274)$&$\chi_{c0}(4500)$&$\chi_{c0}(4700)$\\\hline
Mass&$4146.5$&$4273.3$&$4506$&$4704$ \\
$\sigma_{\rm stat},\sigma_{\rm syst}$&\er$4.5^{+4.6}_{-2.8}$&\er$8.3^{+17.2}_{-\ 3.6}$&\er$11^{+12}_{-15}$&\er$10^{+14}_{-24}$\\\hline
Width&$83$&$56$&$92$&$120$ \\
$\sigma_{\rm stat},\sigma_{\rm syst}$&\er$21^{+21}_{-14}$&\er$11^{+\ 8}_{-11}$&\er$21^{+21}_{-20}$&\er$31^{+42}_{-33}$\\\hline
&$\chi_{c1}(3P)$&$\chi_{c1}(4P)$&$\chi_{c0}(5P)$&$\chi_{c0}(6P)$\\
\hline\hline
\end{tabular}
\end{center}
\renewcommand{\arraystretch}{1.0}
\end{table}

Above this mass, two further $1^{++}$ states were reported, $\chi_{c1}(4140)$ and  $\chi_{c1}(4274)$. 
Both are seen in  $B^\pm\to J/\psi\phi K^\pm$ decays, the former state by several 
collaborations~\cite{Chatrchyan:2013dma,Abazov:2013xda,Lees:2014lra,Aaltonen:2011at,Aaij:2016iza} (only
the latest reference of the collaborations are given here), the latter one by CDF~\cite{Aaltonen:2011at} and
LHCb \cite{Aaij:2016iza}.

The LHCb paper~\cite{Aaij:2016iza} is based on the largest data sample. The amplitude analysis of the reaction 
$B^+\to K^+ \phi J/\psi$ and $\phi\to K^+ K^-$ included the known excited kaon and four
$\omega J/\psi$ resonances. The two lower-mass $\omega J/\psi$ resonances gave the best fit for
$J^{PC}=1^{++}$, the two at higher masses were found to have $J^{PC}=0^{++}$. The results are listed in
Table~\ref{scalar}. The two scalar states are not listed in the PDG summary list.

Isovector states with decay products having hidden charm like $J/\psi$ or
$h_c$ and/or open charm  can obviously not have a pure $c\bar c$ component in their wave function
as they carry exotic quantum numbers. The observations are listed in Table~\ref{XZs}. Three of these states
are accepted by the PDG. Two of them  are $1^{+-}$ states and are therefore called $Z_c({\rm mass})$. 
Most probably, the third accepted state, $X(4020)$ with $I^G=1^+$ has also $J^{PC}=1^{+-}$ quantum numbers.

One state -- seen in its
$\pi\psi(2S)$ decay -- is called $R_{c0}(4240)$.
Its quantum numbers $1^+(0^{--})$ are preferred over $1^+(1^{+-})$ by one standard deviation. This is
presumably insufficient to claim a new resonance, and we combine this observation with $Z_c(4200)$. Masses and widths
are compatible with this identification.  
In the following section
we discuss the implications of their existence from the three different points of view introduced
above.

\section{Discussion}

In this section we now discuss the three views introduced in the introduction in the light of the
mentioned experimental observations.

\subsection{\label{VI}Consequences of View I}

In this view all states near and above the first heavy open flavor $S$-wave threshold with 
matching quantum numbers are classified as molecular states. We begin the discussion with the
charged states. The lowest lying charged states have $J^{PC}=1^{+-}$ ($Z_c(3900)$, $Z_c(4020)$,
$Z_b(10610)$ and $Z_b(10650)$) and are consistent with being molecules formed by
a pseudoscalar and a vector or two vector mesons, respectively.
Moreover, each one of them is located very close to one of the four thresholds,
 $D\bar D^*$, $D^*\bar D^*$, $B\bar B^*$, $B^*\bar B^*$.
A problem occurs with $Z_c(4430)$: This state has the same $J^{PC}$ as the ones mention, however,
it is well above the lowest $S$--wave threshold. Thus, within View I the only possible explanations
for the $Z_c(4430)$ are that it is either a kinematic effect, as proposed in
Ref.~\cite{Pakhlov:2014qva}, or a $P$--wave molecular state composed of $D_1\bar D$ as proposed in
Ref.~\cite{He:2017mbh}.

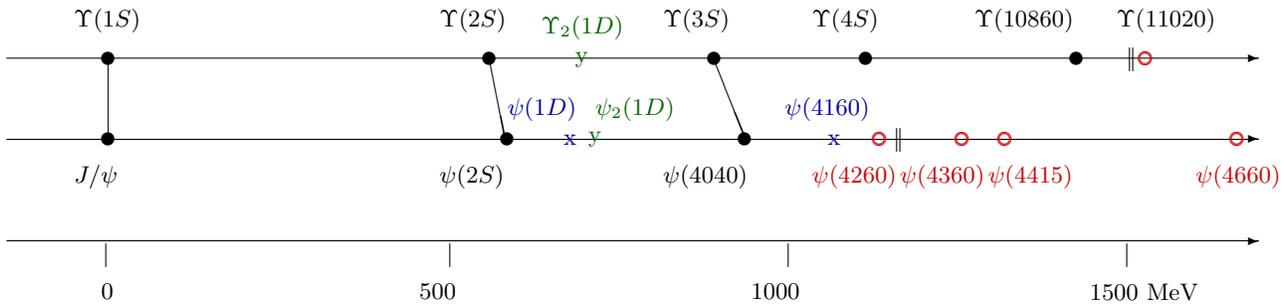
\begin{figure*}[pt]
{\small\vspace{10mm} \setlength{\unitlength}{0.9mm} \hspace*{-140mm}\begin{picture}(10.00,20.00)
\put(-15.00,10.00){\vector(1,0){185.00}} \put(-15.00,22.00){\vector(1,0){185.00}}
 \put(0,22) {\line(0,-1){11}} \put(56.3,22)
{\line(1,-5){2.5}} \put(0.00,10.00){\circle*{2.00}} \put(58.90,10.00){\circle*{2.00}}
\put(94.00,10.00){\circle*{2.00}}
\put(71.00,9.00){\makebox(5.00,2.00)[l]{\gr y}}
\put(132.40,10.00){\rd\circle{2.00}}  \put(132.40,10.00){\rd\circle{1.50}}
\put(67.30,9.00){\makebox(5.00,2.00)[l]{\bl x}}
\put(106.30,9.00){\makebox(5.00,2.00)[l]{\bl x}}
\put(126.10,10.00){\rd\circle{2.00}}\put(126.10,10.00){\rd\circle{1.50}}
\put(0.00,22.00){\circle*{2.00}}
\put(56.30,22.00){\circle*{2.00}} \put(89.50,22.00){\circle*{2.00}}
\put(111.90,22.00){\circle*{2.00}} \put(69.00,20.50){\makebox(5.00,2.50)[l]{\gr y}}
\put(-0.8,-8.00){$\bf\vert$} \put(50.0,-8.00){$\bf\vert$} \put(100.0,-8.00){$\bf\vert$}
\put(150.0,-8.00){$\bf\vert$}
\put(166.70,10.00){\rd\circle{2.00}}\put(166.70,10.00){\rd\circle{1.50}}
\put(113.90,10.00){\rd\circle{2.00}}\put(113.90,10.00){\rd\circle{1.50}}
\put(143.00,22.00){\circle*{2.00}}\put(143.00,22.00){\circle{1.50}}
\put(153.20,22.00){\rd\circle{2.00}}\put(153.20,22.00){\rd\circle{1.50}}
\put(-5.00,25.00){\makebox(15.00,5.00)[l]{$\Upsilon(1S)$}}
\put(49.00,25.00){\makebox(15.00,5.00)[l]{$\Upsilon(2S)$}}
\put(64.00,24.00){\makebox(15.00,5.00)[l]{\gr $\Upsilon_2(1D)$}}
\put(82.00,25.00){\makebox(15.00,5.00)[l]{$\Upsilon(3S)$}}
\put(104.00,25.00){\makebox(15.00,5.00)[l]{$\Upsilon(4S)$}}
\put(128.00,25.00){\makebox(15.00,5.00)[l]{$\Upsilon(10860)$}}
\put(149.00,25.00){\makebox(15.00,5.00)[l]{$\Upsilon(11020)$}}
\put(116.0,9.00){$\bf\vert$}\put(116.5,9.00){$\bf\vert$}
\put(150.4,21.00){$\bf\vert$}\put(150.9,21.00){$\bf\vert$}
\put(89.4,22) {\line(2,-5){4.8}}
\put(-5.00,2.00){\makebox(15.00,5.00)[l]{$J/\psi$}}
\put(49.00,2.00){\makebox(15.00,5.00)[l]{$\psi(2S)$}}
\put(59.00,12.00){\makebox(15.00,5.00)[l]{\bl$\psi(1D)$}}
\put(72.00,12.00){\makebox(15.00,5.00)[l]{\gr$\psi_2(1D)$}}
\put(82.00,2.00){\makebox(15.00,5.00)[l]{$\psi(4040)$}}
\put(100.00,12.00){\makebox(15.00,5.00)[l]{\bl$\psi(4160)$}}
\put(104.00,2.00){\makebox(15.00,5.00)[l]{\rd$\psi(4260)$}}
\put(117.00,2.00){\makebox(15.00,5.00)[l]{\rd$\psi(4360)$}}
\put(130.00,2.00){\makebox(15.00,5.00)[l]{\rd$\psi(4415)$}}
\put(160.70,2.00){\makebox(15.00,5.00)[l]{\rd$\psi(4660)$}}
\put(-1.0,-15.00){\makebox(15.00,5.00)[l]{0}} \put(46.0,-15.00){\makebox(15.00,5.00)[l]{500}}
\put(95.0,-15.00){\makebox(15.00,5.00)[l]{1000}} \put(145.0,-15.00){\makebox(15.00,5.00)[l]{1500
MeV}} \put(-15.00,-5.00){\vector(1,0){185.00}}
\end{picture}\vspace{12mm}
\caption{\label{ccbb1}(Color online) The $\psi_c$ and $\Upsilon$ mass spectra in View I. (Black)
filled circles: $^3S_1$ $Q\bar Q$ mesons;  (blue) x's: $\psi(nD)$  $Q\bar Q$ mesons; 
(green) y's: $D$-wave $Q\bar Q$ mesons; (red) open
circles: extraordinary states. Thresholds for the lowest $S$-wave decays into $B_1\bar B$ or
$D_1\bar D$ are shown by a pair of small vertical bars. 
}}
\end{figure*}
As soon as we accept that the $Z$--states are of molecular nature, we also  expect molecular
states to occur in the isoscalar sector. One argument in support of this is e.g. that for the
exchange of an isovector particle like the pion or the rho meson between two isospin $1/2$ states
one has
\begin{equation}
\label{isospin}
\langle I I_3|\tau_1\cdot \tau_2|I I_3\rangle =\left\{ {\phantom{-}1 \mbox{ for } I=1}\atop{-3 \mbox{ for } I=0}\right .
\ .
\end{equation}
In addition, when the $C$ parity gets switched the central part of the potential acquires an additional sign. Thus, if
isovector exchanges contribute to a relevant amount to the binding of the $Z$--states with  $I=1$ and $J^{PC}=1^{+-}$, then one
should expect that they also generate isoscalar bound states with $J^{PC}=1^{++}$, since the
resulting interaction is attractive in both channels and it is even a factor 3 stronger in the
isoscalar one --- in this sense $Z_c(3900)$ and $\chi_{c1}(3872)$ would be very close relatives.
Accordingly, it should be possible to produce both with an analogous mechanism. This observation
was employed in Ref.~\cite{Guo:2013nza} to predict that $\chi_{c1}(3872)$ must be copiously
produced in $e^+e^-\to\gamma \chi_{c1}(3872)$ given that $Z_c(3900)$ was found in $e^+e^-\to
Z_c(3900)\pi$ --- a prediction confirmed experimentally at BESIII~\cite{Ablikim:2013dyn}
--- these transitions
might well be favored also by a possible molecular structure of the source state for the transition, $\psi(4260)$, (see discussion below)~\cite{Wang:2013cya} and by some kinematical enhancement~\cite{Wang:2013hga}.
. However,
there should not be an isoscalar  $1^{++}$ state near the $D^*\bar D^*$ threshold as a relative of the
$Z_c(4020)$, since $D^*\bar D^*$  in spin 1 $S$-wave has a negative $C$--parity. Moreover,
$if$ single isovector exchanges, like one-pion exchange, provide the most important contribution to the
binding of the $Z$--states via the central potential (that drives the $S$ wave to $S$ wave
transitions), there
should not be any isoscalar bound states with $J^{PC}=1^{+-}$ or isovector ones with
$J^{PC}=1^{++}$ with the same constituents. This is a non-trivial prediction from the molecular
picture. It relies on the additional assumption that single isovector meson exchanges provide an essential
part of the scattering potential that eventually leads to the generation of the molecular states.
Up to date this prediction is in line with observation, however, the recent analysis of
Ref.~\cite{Wang:2018jlv} revealed that the most important contribution
of the one pion exchange contribution comes from the $S-D$ transition driven by the
tensor force. Since this piece enters quadratically into the binding potential in the $S$--wave,
it appears to provide attraction in all channels and the argument presented above should
be refined.

Heavy quark spin symmetry allows one to predict additional states: At leading order heavy quark
spin symmetry one can construct two contact terms for the interaction between the two ground state
$D$--mesons. Moreover, these contact terms appear in the same linear combination in the $1^{++}$
channel and in the $2^{++}$ channel~\cite{Guo:2013sya}. Accordingly, the $1^{++}$ state
$\chi_{c1}(3872)$ located very near the $D\bar D^*$--threshold should have a spin 2 partner
. The latter state
, often called $X_2(4020)$ or -- according to the PDG naming scheme -- $\chi_{c2}(4020)$, is  located close to the
$D^*\bar D^*$ threshold. Once one-pion exchange is included transitions from the $D^*\bar D^*$
$S$-wave with spin 2 to the $D\bar D$-$D$-wave become possible which can lead to widths of a few
MeV~\cite{Albaladejo:2015dsa} to tenth of MeV~\cite{Baru:2016iwj}.

While the data in the isoscalar states in the charmonium mass range is insufficient to fix the two
mentioned contact terms the situation is different for the isovector states. For example the heavy
quark spin partners of the $Z_b$ states were studied in
Refs.~\cite{Voloshin:2011qa,Mehen:2011yh,Guo:2013sya,Baru:2019xnh}.

The interpretation of $\chi_{c1}(3872)$ as $D^*\bar D$ molecule suggests the existence of a 
$D_s^*\bar D_s$ molecule close to 4.081\,GeV. 
However, the mass of the next heavier state with $J^{PC}=1^{++}$ is $\chi_{c1}(4140)$ at $4146.8\pm 2.4$\,MeV.
It is thus located about 60\,MeV above the mentioned threshold. Thus a $D_s^*\bar D_s$
molecular nature of this state is difficult to anticipate in particular since it would then predominantly
decay into that channel which should lead to a significantly broader width than the observed 20\,MeV.
An alternative possibility could be a $D_s^*\bar D_s^*$   bound state whose threshold is located 80\,MeV
above the $\chi_{c1}(4140)$. However, so far basically all calculations could get a state
in this mass range from $D_s^*\bar D_s^*$ only with $J^{PC}=0^{++}$ or $2^{++}$ ---
for a recent summary of the situation we refer to Ref.~\cite{Wang:2017mrt}. In this work
the structure called $X(4160)$ in the RPP is claimed to originate from the $D_s^*\bar D_s^*$
dynamics.
As it stands, also the  $\chi_{c1}(4274)$ -- if confirmed -- seems incompatible with View I.

There is an interesting difference between the molecular picture and the predictions of the quark
model already for the $D^{(*)}\bar D^{(*)}$ systems: While in the quark model there exists for each
radial excitation one state with $J^{PC}=1^{+-}$, in the molecular picture there are typically two,
since both the $D\bar D^*$ as well as the $D^*\bar D^*$ channel can couple to these quantum numbers
in an $S$--wave. This is the pattern already seen for the charged states, however, so far not in
the isoscalar channels.

Besides  non--perturbative $D\bar D$, $D^*\bar D$ and $D^{*}\bar D^{*}$ interactions, also non--perturbative
interactions between $D_{1}$ or $D_2^*$ and $\bar D$ or $\bar D^*$ mesons
can lead to hadronic molecular states.
Since the $(D_1,D_2^*)$ spin multiplet carries even parity, those will have odd parity for $S$--wave
interactions. In View I one is thus obliged to interpret the $1^{--}$ states $\psi(4260)$ aka
$Y(4260)$, $\psi(4360)$ aka $Y(4360)$ as well as $\psi(4415)$ as $D_1\bar D$, $D_1\bar D^*$ and
$D_2^*D^*$ molecular states, respectively~\cite{Wang:2013cya,Cleven:2015era,Wang:2013kra}. This nicely explains
why negative parity exotic candidates  are about 400 MeV heavier that their even parity partners:
This mass difference reflects the $D_1(2420)$--$D^*$ mass difference. Accordingly, the molecular
picture predicts that exotic pseudoscalars should be even heavier, since the quantum numbers
$0^{-+}$ can only be reached in an $S$-wave with the constituents  $D_1\bar D^*$, with a threshold
yet another 140 MeV higher~\cite{Cleven:2015era}. Moreover, one expects a clear signal of $Y(4260)$
in the $D\bar D^*\pi$ final state since $D^*\pi$ is the main decay channel of the $D_1$. This
signal was predicted in Ref.~\cite{Cleven:2013mka} and confirmed recently at
BESIII~\cite{Ablikim:2018vxx}.

For the vector states, view I is summarized in Fig.~\ref{ccbb1}. The $Q\bar Q$ mesons are
represented by (black) filled circles or, for states assigned to $D$-waves, by (green) crosses, the
extraordinary states by (red) open circles.  The thresholds for the lowest $S$-wave decays into
$B_1\bar B$ or $D_1\bar D$ are shown by a pair of small vertical bars. One state below this
threshold and all states above are classified as extraordinary states. In this view the $\psi(4660)$ aka $Y(4660)$
is to be seen as $\psi(2S)f_0(980)$ molecular state~\cite{Abe:2007jna}, since it is located
right at the corresponding threshold. A non-trivial prediction that emerges from this assignment
is the existence of a spin partner of the $Y(4660)$ --- a bound state of $\eta_c(2S)$ and $f_0(980)$~\cite{Guo:2009id}.

Unfortunately within a hadronic effective field theory it is not possible to make the relation between the
molecular structures in the bottomonium sector and those in the charmonium sector quantitative~\cite{Baru:2018qkb}.
It therefore appears not possible to predict the spectrum of exotic bottomonia from the rich spectrum of
charmonia.

A very interesting aspect that comes with the molecular assignment is the role of the heavy quark
spin in decays: In the heavy quark limit the spin of the heavy quark decouples from the
system. In addition, in $e^+e^-$ collisions that lead to final states which contain a heavy
quark and its antiquark, the heavy quark-antiquark pair is produced directly at the photon
vertex in spin 1 to avoid the need to generate heavy quarks via gluonic interactions --- a process
suppressed as a result of asymptotic freedom.
Accordingly one would expect to see in reactions like $e^+e^-\to (\bar QQ)2\pi$ the $\bar QQ$ pair
by far predominantly in a spin one state. For the low lying heavy quarkonia this is indeed the
case --- for example the branching ratio $\Upsilon(3S)\to \Upsilon(2S)\pi^+\pi^-$ is with 4.4\%
more than two orders of magnitude larger than the upper bound for $\Upsilon(3S)\to h_b(1P)\pi^+\pi^-$
currently quoted as $1.2\times 10^{-4}$.
In contrast to this the $Z_b$ states decay equally often into spin 1 and spin 0 final states.
This pattern finds a natural explanation in the molecular picture where
~\cite{Bondar:2011ev}
\begin{eqnarray} \nonumber
Z_b(10610)\sim B^*\bar B-B\bar B^* =\frac{1}{\sqrt{2}}\left(0^-_{\bar qq}\otimes 1^-_{\bar bb} - 1^-_{\bar qq}\otimes 0^-_{\bar bb}\right), &&\\
Z_b(10650)\sim \quad~B^*\bar B^*\quad~=\frac{1}{\sqrt{2}}\left(0^-_{\bar qq}\otimes 1^-_{\bar bb} + 1^-_{\bar qq}\otimes 0^-_{\bar bb}\right).&&
\end{eqnarray}
The $Z_b$ states were observed in the decays of $\Upsilon(10860)$ which is far way from an $S$--wave threshold with
matching quantum numbers.
Interestingly the data in the $h_b\pi$ final states accordingly are non-vanishing only near the $Z_b$ peaks, for apparently in
this case the $Z_b$ intermediate states are necessary to populate the final state with heavy quark spin equal to 0.
This picture is supported by the fact that in the related $\Upsilon(nS)\pi$ spectra there is a lot of strength even outside
the $Z_b$ peaks.
The situation is very much different in case of the decays of $Y(4260)$: While this state is also seen in both
$h_c\pi\pi$ and $J/\psi\pi\pi$ final states with similar strength, here even in the data with an $h_c$ in the
final state there is a lot of strength observed also outside the $Z_c$ peaks. This could be interpreted
as an indication of a $D_1\bar D$ molecular nature of the $Y(4260)$, since~\cite{Li:2013yka}
\begin{equation}
(D_1\bar D - \bar D_1 D) \sim \frac{1}{2\sqrt{2}} \psi_{11}+\frac{\sqrt{5}}{2\sqrt{2}} \psi_{12}+\frac{1}{2} \psi_{01} ,
\end{equation}
where $\psi_{1J}=1_H^{--}\otimes J_L^{++}$ and $\psi_{01}=0_H^{-+}\otimes 1_L^{+-}$  where
$H$ and $L$ stand for the heavy and light quark-antiquark pair. The analogous argument
applies for $Y(4360)$. The presence of light quarks in the wave function for
$Y(4260)$ finds further support in the analysis of Ref.~\cite{Chen:2019mgp}.

\subsection{Consequences of View II}

We start the discussion of View II with $\chi_{c1}(3872)$ aka $X(3872)$. In this view, this state is identified with $\chi_{c1}(2P)$.
The view does not deny that $X(3872)$ has a large $D^*\bar D$ component: its mass happens to be close
to the sum of the $D^{*0}$ and $\bar D^{0}$ masses and is synchronized to match
the exact mass~\cite{Bugg:2008wu}. The $c\bar c$ pair creates a $d\bar d$ pair thus
dressing the $c\bar c$ core with a $D^{*0}\bar D^{0}$ cloud. The pole leads to a strong $S$-wave
attraction between the $D^{*0}$ 
and $\bar D^{0}$ mesons. However, only one resonance is formed. 

One might think that
a hard $c\bar c$ core of $\chi_{c1}(3872)$ is needed to explain its copious  production 
in hard processes in $\bar pp$ collisions
at $\sqrt s$=1.96\,TeV~\cite{Acosta:2003zx} and at 7\,TeV~\cite{Aaij:2011sn}.
This reasoning is put forward in Refs.~\cite{Bignamini:2009sk,Bignamini:2009fn,Esposito:2015fsa,Esposito:2017qef}
but was criticised in Refs.~\cite{Artoisenet:2009wk,Artoisenet:2010uu,Albaladejo:2017blx,Guo:2014sca,Guo:2014ppa,Braaten:2018eov}.
At this point in time it is fair to say that there is no consenus in the literature wether or
not large momentum transfer reactions can act as a filter to states with quark--anti-quark cores.

Using QCD sum rules to calculate the width of the 
radiative decay of the $\chi_{c1}(3872)$ meson, the authors of Ref.~\cite{Matheus:2009vq,Nielsen:2010ij}
conclude that the X(3872) is approximately 97\% a charmonium state with a 3\% admixture of 
$D^0D^{*0}$ molecule ($\sim 88$\%) and a $D^+D^{*-}$ molecule ($\sim 12$\%). 
This result is basically reproduced in Ref.~\cite{Zanetti:2011ju};  the authors discuss the intrinsic
uncertainties in their calculations and underline that the $c\bar c$ part of the state plays a very important role
in the determination of branching ratios. The authors of 
Ref.~\cite{Meng:2014ota} study the $B\to K D\bar D^*$ and $B\to K J/\psi\pi^+\pi^-$ decay processes
and claim that  $X(3872)$ is mainly a $c\bar c$ resonance with a small contribution generated by 
$\bar DD^*$ final state interaction. In \cite{Zhou:2017dwj,Zhou:2017txt}, the 
$\chi_{c1}(3872)$ is estimated to have a strong
molecular and a 10-30\% $c\bar c$ component. From a study of radiative decays, the authors of 
Ref.~\cite{Cincioglu:2019gzd} conclude that a wide range of $c\bar c$ versus molecular components is 
consistent with the data. However, in Ref.~\cite{Guo:2014taa} it is pointed out  that it is impossible to quantify the 
impact of the short-range contributions in an effective field theory based on hadrons and radiative 
decays of the $\chi_{c1}(3872)$ are sensitive to this part of the wave function. The radiative decays can hence
not be used as an argument disfavoring the molecular view but the molecular view can -- based on the same argument -- 
not rule out a significant $Q\bar Q$ core.


The next state with $J^{PC}=1^{++}$ above $\chi_{c1}(3872)$ is 
$\chi_{c1}(4140)$ which is seen only in its decay into $J/\psi\phi$.
In View II, it is interpreted as $\chi_{c1}(3P)$,  followed by $\chi_{c1}(4274)$ as $\chi_{c1}(4P)$.
In View I,  $\chi_{c1}(4274)$ can possibly be understood as a bound state of 
$D_s^*\bar D_s^*$~\cite{Branz:2009yt}. In View II, it has a $c\bar c$ core but
could be dressed by a $D_{s}^*\bar D_{s}$  cloud.

View II identifies the $\chi_{c2}(3930)$ state with $\chi_{c2}(2P)$. 
Likewise, the $\chi_{c0}(3860)$ candidate is interpreted as $\chi_{c0}(2P)$ state. Both resonances fall just in 
between important thresholds at 3730, 3972 and 4014\,MeV. 

\begin{figure*}[pt]
{\small\vspace{10mm} \setlength{\unitlength}{0.9mm} \hspace*{-140mm}\begin{picture}(10.00,20.00)
\put(-15.00,10.00){\vector(1,0){185.00}} \put(-15.00,22.20){\vector(1,0){185.00}}
\put(0.00,10.00){\circle*{2.00}}
\put(0,22) {\line(0,-1){11}}
 \put(58.90,10.00){\circle*{2.00}}
\put(56.3,22) {\line(1,-5){2.5}} \put(94.10,10.00){\circle*{2.00}}
\put(113.30,10.00){\rd\circle*{0.80}} \put(113.30,10.00){\rd\circle{2.00}}
\put(72.50,8.60){\makebox(5.00,2.50)[l]{\gr y}} \put(132.40,10.00){\circle*{2.00}}
\put(166.70,10.00){\circle*{2.00}} \put(166.70,10.00){\circle{1.50}}
\put(68.60,8.60){\makebox(5.00,2.50)[l]{\bl x}}
\put(109.40,8.60){\makebox(5.00,2.50)[l]{\bl x}} \put(127.10,8.60){\makebox(5.00,2.50)[l]{\rd
\circle*{0.80}}} \put(127.10,8.60){\makebox(5.00,2.50)[l]{\rd \circle{2.00}}}
\put(0.00,22.00){\circle*{2.00}} \put(56.30,22.00){\circle*{2.00}}
\put(89.50,22.00){\circle*{2.00}} \put(89.4,22) {\line(2,-5){4.8}}
\put(111.90,22.00){\circle*{2.00}} \put(111.9,22) {\line(5,-3){20}}
\put(143.00,22.00){\circle*{2.00}} \put(143.00,22.00){\circle{1.50}}
\put(143.0,22){\line(2,-1){24}} \put(153.20,22.00){\circle{2.00}}
\put(69.00,21.00){\makebox(5.00,2.50)[l]{\gr y}} \put(-0.8,-8.00){$\bf\vert$}
\put(50.0,-8.00){$\bf\vert$} \put(100.0,-8.00){$\bf\vert$} \put(150.0,-8.00){$\bf\vert$}
\put(-5.00,25.00){\makebox(15.00,5.00)[l]{$\Upsilon(1S)$}}
\put(52.00,25.00){\makebox(15.00,5.00)[l]{$\Upsilon(2S)$}}
\put(64.00,24.00){\makebox(15.00,5.00)[l]{\gr $\Upsilon_2(1D)$}}
\put(85.00,25.00){\makebox(15.00,5.00)[l]{$\Upsilon(3S)$}}
\put(106.00,25.00){\makebox(15.00,5.00)[l]{$\Upsilon(4S)$}}
\put(132.00,25.00){\makebox(15.00,5.00)[l]{$\Upsilon(10860)$}}
\put(149.00,25.00){\makebox(15.00,5.00)[l]{$\Upsilon(11020)$}}
\put(-5.00,2.00){\makebox(15.00,5.00)[l]{$J/\psi$}}
\put(55.00,2.00){\makebox(15.00,5.00)[l]{$\psi(2S)$}}
\put(62.00,12.00){\makebox(15.00,5.00)[l]{\bl$\psi(1D)$}}
\put(72.00,12.00){\makebox(15.00,5.00)[l]{\gr$\psi_2(1D)$}}
\put(86.00,2.00){\makebox(15.00,5.00)[l]{$\psi(4040)$}}
\put(87.00,-2.50){\makebox(15.00,5.00)[l]{$\psi(3S)$}}
\put(96.00,12.00){\makebox(15.00,5.00)[l]{\bl$\psi(4160)$=$\psi(2D)$}}
\put(103.00,2.00){\makebox(15.00,5.00)[l]{\rd$\psi(4260)$}}
\put(115.50,2.00){\makebox(15.00,5.00)[l]{\rd$\psi(4360)$}}
\put(128.70,2.00){\makebox(15.00,5.00)[l]{$\psi(4415)$}}
\put(129.70,-2.50){\makebox(15.00,5.00)[l]{$\psi(4S)$}}
\put(160.70,2.00){\makebox(15.00,5.00)[l]{$\psi(4660)$}}
\put(-1.0,-15.00){\makebox(15.00,5.00)[l]{0}} \put(46.0,-15.00){\makebox(15.00,5.00)[l]{500}}
\put(95.0,-15.00){\makebox(15.00,5.00)[l]{1000}} \put(145.0,-15.00){\makebox(15.00,5.00)[l]{1500
MeV}} \put(-15.00,-5.00){\vector(1,0){185.00}}
\end{picture}\vspace{12mm}
\caption{\label{ccbb2}The $\psi_c$ and $\Upsilon$ mass spectra in View II.
The $\psi(3770)$ and $\psi(4160)$ are interpreted as $\psi(1D)$ and $\psi(2D)$ and marked by (blue) x's, 
the two states with $J^{PC}=2^{--}$ by (green) y's. The
$\psi(4040)$ and $\psi(4415)$ states are interpreted 
as $\psi(3S)$ and $\psi(4S)$. The $\psi(4260)$ and $\psi(4360)$, mostly interpreted as {\it unconventional states}, are
suggested to be generated by an interplay of interference effects and threshold singularities.
At this stage the  $\Upsilon(11020)$ does not fit into the scheme.
Note that the PDG mass values sometimes differ significantly
from the mass given in the name of a state. \vspace{2mm}
}}
{\small\vspace{10mm} \setlength{\unitlength}{0.9mm} \hspace*{-140mm}\begin{picture}(10.00,20.00)
\put(-15.00,10.00){\vector(1,0){185.00}} \put(-15.00,22.00){\vector(1,0){185.00}}
 \put(0,22) {\line(0,-1){11}} \put(56.3,22)
{\line(1,-5){2.5}} \put(0.00,10.00){\circle*{2.00}} \put(58.90,10.00){\circle*{2.00}}
\put(94.00,10.00){\circle*{2.00}}
\put(71.00,9.00){\makebox(5.00,2.50)[l]{\gr y}} \put(132.60,10.00){\circle*{2.00}}
\put(67.30,9.00){\makebox(5.00,2.50)[l]{\bl x}}
\put(106.30,9.00){\makebox(5.00,2.50)[l]{\bl x}}
\put(126.10,8.60){\makebox(5.00,2.50)[l]{\bl x}}
\put(0.00,22.00){\circle*{2.00}}
\put(56.30,22.00){\circle*{2.00}} \put(89.50,22.00){\circle*{2.00}}
\put(111.90,22.00){\circle*{2.00}} \put(69.00,21.00){\makebox(5.00,2.50)[l]{\gr y}}
\put(-0.8,-8.00){$\bf\vert$} \put(50.0,-8.00){$\bf\vert$} \put(100.0,-8.00){$\bf\vert$}
\put(150.0,-8.00){$\bf\vert$} \put(166.70,10.00){\circle{2.00}}
\put(113.90,10.00){\circle*{2.00}}
\put(142.80,22.00){\circle*{2.00}}
\put(153.20,22.00){\circle{2.00}}
\put(-5.00,25.00){\makebox(15.00,5.00)[l]{$\Upsilon(1S)$}}
\put(49.00,25.00){\makebox(15.00,5.00)[l]{$\Upsilon(2S)$}}
\put(64.00,24.00){\makebox(15.00,5.00)[l]{\gr $\Upsilon_2(1D)$}}
\put(82.00,25.00){\makebox(15.00,5.00)[l]{$\Upsilon(3S)$}}
\put(104.00,25.00){\makebox(15.00,5.00)[l]{$\Upsilon(4S)$}}
\put(128.00,25.00){\makebox(15.00,5.00)[l]{$\Upsilon(10860)$}}
\put(149.00,25.00){\makebox(15.00,5.00)[l]{$\Upsilon(11020)$}}
\put(89.4,22) {\line(2,-5){4.8}}
\put(111.7,22) {\line(1,-5){2.5}}
\put(142.8,22){\line(-4,-5){10}}
\put(-5.00,2.00){\makebox(15.00,5.00)[l]{$J/\psi$}}
\put(55.00,2.00){\makebox(15.00,5.00)[l]{$\psi(2S)$}}
\put(60.00,12.00){\makebox(15.00,5.00)[l]{\bl$\psi(1D)$}}
\put(72.00,12.00){\makebox(15.00,5.00)[l]{\gr$\psi_2(1D)$}}
\put(88.00,2.00){\makebox(15.00,5.00)[l]{$\psi(4040)$}}
\put(90.00,-2.50){\makebox(15.00,5.00)[l]{$\psi(3S)$}}
\put(96.00,12.00){\makebox(15.00,5.00)[l]{\bl$\psi(4160)$}}
\put(108.00,2.00){\makebox(15.00,5.00)[l]{$Y(4260)$}}
\put(109.70,-2.50){\makebox(15.00,5.00)[l]{$\psi(4S)$}}
\put(117.00,12.00){\makebox(15.00,5.00)[l]{\bl$\psi(4360)$}}
\put(127.70,2.00){\makebox(15.00,5.00)[l]{$\psi(4415)$}}
\put(129.70,-2.50){\makebox(15.00,5.00)[l]{$\psi(5S)$}}
\put(150.70,2.00){\makebox(15.00,5.00)[l]{$\psi(4660)$}}
\put(-1.0,-15.00){\makebox(15.00,5.00)[l]{0}} \put(46.0,-15.00){\makebox(15.00,5.00)[l]{500}}
\put(95.0,-15.00){\makebox(15.00,5.00)[l]{1000}} \put(145.0,-15.00){\makebox(15.00,5.00)[l]{1500
MeV}} \put(-15.00,-5.00){\vector(1,0){185.00}}
\end{picture}\vspace{12mm}
\caption{\label{oldview}The $\psi_c$ and $\Upsilon$ mass spectra. The $\psi(4040)$, $\psi(4260)$, and
$\psi(4415)$ have masses which suggest that there should be identified with $\psi(3S)$ to
$\psi(5S)$, they are marked by filled circles. The $\psi(4160)$ and $\psi(4360)$ are
marked with (blue) x's and identified with $\psi(2D)$ and $\psi(3D)$. $J^{PC}=2^{--}$ states
are shown by (green) y's. The identification of $\psi(4660)$ and $\Upsilon(11020)$ is open.}}
\end{figure*}

If $X(3872)$ can be identified with $\chi_{c1}(2P)$, we can apply the center-of-gravity rule to predict
the mass of the $h_c(2P)$. The center-of-gravity rule of Eq.~(\ref{cog}) proved to be
well satisfied for $b\bar b$ mesons in the $1P$ and $2P$ level and for $c\bar c(1P)$ mesons.  
In Ref.~\cite{Lebed:2017yme}
it is argued that the center-of-gravity rule~of Eq.~(\ref{cog}) can be used as a diagnostic for $\bar QQ$ states.
Since only three states in the $2P$ level are known, $\chi_{c2}(3930)$, $\chi_{c1}(3872)$, $\chi_{c0}(3860)$, we
can use Eq.~(\ref{cog}) only to predict the $h_{c}(2P)$ mass and find
 \begin{eqnarray}
M_{h_{c}(2P)}  &=&    (3900\pm 10){\rm \,MeV.}
\end{eqnarray}

In View II, we thus require the existence of one and only one $1^{+-}$ state at a mass of about
3900\,MeV.  This is in line with Ref.~\cite{Zhou:2017dwj,Zhou:2017txt}: The authors identify
$\chi_{c0}(3862)$, $\chi_{c1}(3872)$, and $\chi_{c2}(3930)$ with the $\chi_c(2P)$ states and 
calculate the $h_c(2P)$ mass to 3902\,MeV. 

This is clearly a different prediction compared to what emerges in the molecular picture of View I, where,
as mentioned above, two $1^{+-}$ can (but don't need to) emerge, one near to the $D\bar D^*$ threshold (and
thus close to 3900 MeV) and one close to the $D^*\bar D^*$ threshold at 4020 \,MeV. 

Mass, production and decay modes of $X(3940)$ are compatible with an assignment to $\eta_c(3S)$ but
its quantum numbers have not yet been determined. As a pseudoscalar state, it is compatible with View I and
View II, since the lowest relevant threshold is much higher than 3940 MeV.

Now we turn to the vector states. 
Figures~\ref{ccbb2} and \ref{oldview} compare the spectrum of $\Upsilon$ states with that of $\psi$ states
in two different realisations of View II. First we discuss their common features.

$\Upsilon(2S)$ has a mass 563\,MeV above $\Upsilon(1S)$, $\psi(2S)$ is found 589\,MeV above
$J/\psi$. $\Upsilon(3S)$ is situated 331.5\,MeV above $\Upsilon(2S)$. We thus expect the $\psi(3S)$
state just above 4035\,MeV. Indeed, a $\psi(4040)$ exists. In the quark model, we interpret this
state, due to its mass and decay modes, as $\psi(3S)$. 
The quantum numbers $I^G(J^{PC})=0^-(2^{--})$ were suggested for
 $\psi_2(3823)$ and $\Upsilon_2(1D)$, but both need
confirmation. If confirmed they seem to be related and $\psi_2(3823)$ is likely $\psi_2(1D)$. The
$\psi(3770)$ state is close in mass to $\psi_2(3883)$; in the quark model it can be assigned
to the $\psi(1D)$ state, likely with some mixing with $\psi(2S)$. 
Above this mass, the two scenarios differ. The first scenario assumes that
$\psi(4260)$ and $\psi(4360)$ are not real resonances, the second scenario assigns the two resonances
to the series of $\psi(nS)$ and $\psi(nD)$ resonances. In both scenarios $\psi(4415)$ is assigned to
the $\psi(nS)$ series and interpreted as $\psi(4S)$ in the first, as $\psi(5S)$ in the second scenario.

\underline{First scenario:} 
$\Upsilon(4S)$ is situated 224\,MeV above $\Upsilon(3S)$. We may thus expect $\psi(4S)$ at about
4300\,MeV. There are claims  for four states close in mass, see Table~\ref{satellites}. Their masses
could be acceptable, however, as states with a large $Q\bar Q$ component at the origin, one may
expect that their $e^+e^-$ decay width should be larger; the decay modes of these states are completely
unexpected. Possibly, none of them is the $\psi(4S)$. Hence we first try to identify 
$\psi(4415)$ with $\psi(4S)$.

Table~\ref{psi} collects some results on vector charmonium states. In an R scan above $\psi(2S)$, 
the BES collaboration observed four further states, $\psi(3770)$, $\psi(4040)$, $\psi(4160)$,
$\psi(4415)$~\cite{Ablikim:2007gd}. While $\psi(3770)$ decays predominantly into $D\bar D$,
the other mesons, including $\psi(4260)$ and $\psi(4360)$, have a large number of decay options. 
However, none of them listed by the PDG with a finite branching ratio, except of the 
(10\er4)\% for the $\psi(4415)\to D_2^*\bar D$ decay~\cite{Pakhlova:2007fq}.
Apart from the ``stable'' $J/\psi$ and
$\psi(2S)$, the total widths of these states have a similar magnitude. The $e^+e^-$ partial decay
widths decrease steadily for the states assigned in Table~\ref{psi} to the $nS$ series; states assigned to the $nD$
series have somewhat smaller $e^+e^-$ partial decay widths: The density of the wave function near the
origin decreases with $n$ and is smaller for $D$-wave states. The relatively large value for
$\psi(4160)$ may indicate a significant $3S$ contribution.

\begin{table}[pt]
\renewcommand{\arraystretch}{1.2}
\caption{\label{psi}Charmonium states with $J^{PC}=1^{--}$ in the first scenario of View II.}
\begin{tabular}{lccclcr}
\hline\hline\vspace{-3mm}\\
State & \multicolumn{2}{c}{$\Gamma_{\rm tot}$}&\quad&\multicolumn{2}{c}{
$\Gamma_{e^+e^-}$} & Assigned \\[1ex]
\hline
$J/\psi$        &$0.0929$$\pm0.0028$&MeV&&5547&eV&$\psi(1S)$\\
$\psi(2S)$     &$0.294$$\pm$$0.008$&MeV&&2331&eV&$\psi(2S)$ \\
$\psi(3770)$ &$87.04$$\pm$$0.35$  &MeV&&261&eV&$\psi(1D)$ \\
$\psi(4040)$ &$80\pm10$                &MeV&& 850&eV&$\psi(3S)$ \\
$\psi(4160)$ &$70\pm10$                &MeV&& 483&eV&$\psi(2D)$ \\
$\psi(4415)$ &$62\pm20$                &MeV&& 580&eV&$\psi(4S)$ \\\hline\hline
\end{tabular}
\renewcommand{\arraystretch}{1.}\end{table}

The $\psi(4415)$ fits into this series, and based on Table~\ref{psi}, we are tempted to
interpret this state as $Q\bar Q$. When its mass is compared to $\Upsilon(4S)$, it seems to have too
high a mass. On the other hand, its mass is just on the $D_{s1}\bar D_{s}$ threshold and close to the $D_2\bar D^*$ threshold. A threshold is known to
attract the pole position~\cite{Bugg:2008wu} --- the possible impact of the hidden strangeness
channel on the $\psi(4415)$ is discussed in Ref.~\cite{Cao:2017lui}. 
Hence it is not unreasonable to identify $\psi(4415)$ with $\psi(4S)$.
Due to its mass at the $D_{s1}\bar D_s$  threshold it  
develops a significant molecular component~\cite{Wang:2013cya,Cleven:2015era} which dominates the
decay modes. However, in View II it owes its existence to its $Q\bar Q$ core.

There are eight observations of resonances with identical quantum numbers 
claimed in the mass interval between 4200 and 4400 MeV.
Even if we combine five of these observations to two states (see Table~\ref{satellites}), the density
of states is rather high.
The peak cross sections for $\psi(4160)$ and $\psi(4415)$ are about 15.000\,pb, with
significant interference between the two states~\cite{Ablikim:2016qzw}. The cross section for
$e^+e^-\to\psi(4260)$ and $\psi(4360)\to J/\psi \pi^+\pi^-$ is  with at most 80\,pb less than 1\% of the
cross section for $\psi(4160)$ and $\psi(4415)$.
A possible explanation for the structures called $\psi(4260)$ and $\psi(4360)$
within View II could be that the two resonances $\psi(4160)$ and $\psi(4415)$ both have
a small coupling to OZI violating final states. In Ref.~\cite{Chen:2017uof} it is
shown how interference effects can then lead to the observed structures in the absence of true
resonances. The authors of Ref.~\cite{Chen:2017uof} suggest that only $\psi(4220)$ might survive as
a resonance. However, the best estimate
for the mass of $\psi(4160)$ is 4190\,MeV, just 30\,MeV lower than the mass of the conjectured
$\psi(4220)$. These two observations may hence be one single state.

There is an important caveat: The fits of Ref.~\cite{Chen:2017uof} reveal that the existing
data in the $h_c\pi\pi$ and $J/\psi\pi\pi$ channels can be described 
only, if one allows for a huge violation of spin symmetry which appears not to be natural.
Moreover, the analysis of Ref.~\cite{Chen:2017uof} introduces a strong non-resonant background,
also with large spin symmetry violation and fine tuned in each channel to suppress the true
resonance signals at the resonance locations. Nevertheless, the analysis shows that there might be solutions in which
the states between $\psi(4160)$ and $\psi(4415)$ are not needed. We remind the reader that there are
very significant thresholds between 4100 and 4500\,MeV: $S$-wave decays into $D_1\bar D$ are open at 4286\,MeV,
into $D_{s1}\bar D_s$ at 4428\,MeV; $P$ wave decays into $D_s^*\bar D_s$ open at
4.081\,MeV, into $D_1\bar D^*$ at 4228\,MeV. A coupled-channel analysis of all final states with proper evaluation
of all kind of singularities is missing. Possibly, there is no state at all between
$\psi(4190)$ and $\psi(4415)$. 

\underline{Second scenario:}
An alternative interpretation of the $\psi$ mass spectrum within the $Q\Bar Q$ quark model
was given in Ref.~\cite{Klempt:2007cp} where 
the $\psi(4260)$ was interpreted as $\psi(4S)$, the $\psi(4360)$ as $\psi(3D)$ and $\psi(4415)$ as
$\psi(5S)$. Figure~\ref{oldview} shows that interpretation. 
The excitation energies of the low-mass charmonium system are increasingly 
larger than those in the bottomonium spectrum, while the opposite pattern is observed at higher masses.
This looks unnatural.  We emphasize, however, that the identification of
 $\psi(4260)$ with $\psi(4S)$, of $\psi(4360)$ with $\psi(3D)$ and of  $\psi(4415)$ with
$\psi(5S)$ is an over-simplification. All states can mix: Ref.~\cite{Wang:2018rjg} gives a fit to 
data assuming that the $4S$ and $3D$ states mix to form $\psi(4260)$ and $\psi(4360)$
and the $5S$ and $4D$ mix to form the $\psi(4415)$ and a hypothetical $\psi(4500)$. Further, all
states may contain a molecular component. View II just claims that these states have a $c\bar c$ seed
and that this seed is decisive for the existence of these states. 

The first scenario is approximately compatible with quark-model calculations using the Cornell potential;
the second scenario is compatible when a screened potential is used.  This is discussed in Section~\ref{Models}.

In quark models with $Q\bar Q$ states only, charged heavy-quark states are not admitted.  
In View II the structures seen in isovector channels are not regular resonances
with a pole in the $S$--matrix. The $Z_c$ and $Z_b$ states would
have to be produced via perturbative rescattering in the final state. At present, this possibility is not
(yet?) ruled out. A softer version of View II forbids only
{\it hidden exotics} but allows for states beyond the quark model which cannot be reduced
to $Q\bar Q$, which means that they  have quantum numbers not accessible to $Q\bar Q$.

The $\Upsilon(10860)$ could decay into $B^*\bar B\pi$ where the  $B^*\bar B$ system
rescatters into $\Upsilon(nS)\pi$ when $B^*$ and $\bar B$ are in $S$-wave and have little 
relative momentum.  It was shown in Ref.~\cite{Guo:2015umn} that rescattering can
produce a loop in an Argand diagram. In various works, several of the $XYZ$ states are suggested to
be only a kinematical
enhancement~\cite{Bugg:2004rk,Bugg:2011jr,Chen:2011pv,Chen:2011xk,Chen:2011pu,Chen:2013coa,Chen:2013wca,Swanson:2014tra,Swanson:2015bsa}.
A newly created $c\bar c$ pair
has a strong affinity to decay into a pair of (possibly excited) $D$ or $D_s$ mesons. There are a
few $D/D_s$ mesons which are narrow, so that rescattering is possible. The thresholds for the
production of a pair of narrow charmed mesons are given in Table~\ref{thresholds}. The $X(4100)$
has no close-by threshold but it is seen only as effect with three standard deviation significance.
Also $Z_{c}(4430)$ has no close-by $S$-wave threshold but is seen as a clear signal. There are two
thresholds close to its mass, for $D_1\bar D^*$ and $D_{s1}\bar D_s$ but in $S$-wave they belong to
$0^{-+}$ and $1^{--}$ quantum numbers. Still, as mentioned above, this state could have emerged
from a triangle singularity~\cite{Pakhlov:2014qva}.

The authors of Ref.~\cite{Pilloni:2016obd} study possible scenarios for the related $Z_c(3900)$
which might be a compact QCD state, a virtual state, or a kinematical enhancement. 
The authors conclude
that current data are not precise enough to distinguish between these hypotheses. In
Ref.~\cite{Guo:2014iya} it was demonstrated that in particular the channel that is related to the
nearby threshold is sensitive to the presence or the absence of a pole. Thus here, experiment will
eventually be able to tell the difference. The reasoning of Ref.~\cite{Guo:2014iya} was questioned
in Ref.~\cite{Swanson:2015bsa}. There, however, different physics drives the structures in the
near-by channel and the others: Formfactors in the former, cusps in the latter. This is not only
unsatisfying, it also leads to the prediction that there should be similar structures near each
$S$-wave threshold which does not seem to be the case.
But given the controversy in the literature one may conclude that at present
there is no
forcing evidence that genuine isovector states with a normal pole structure in the complex
scattering plane exist.

\section{Consequences of View III}

In the previous sections we have seen that both View I and View II have problems with certain states in the
spectrum. Some of those problems might get resolved with the appearance of better data, but some might
not. We would therefore like to now confront the current data situation with View III, which allows for
the co-existence of molecular states and quark model states. 
In the literature there are various model calculations that try to capture this view --- see, e.g., Ref.~\cite{Ortega:2019tby}
and references therein. 

To get a physics understanding of how such an interplay of different structures could emerge in QCD
we could start from the large $N_c$ limit of QCD~\cite{tHooft:1973alw}. In this limit there exist infinite
towers of stable  $\bar QQ$ states (and eventually also tetraquarks which were added to
the discussion in Ref.~\cite{Weinberg:2013cfa}). As one then starts reducing the number of 
colors gradually, the coupling of the quark model states to two--meson states grows.
Accordingly the states above the two hadron--threshold acquire a finite life time.
It was demonstrated in Ref.~\cite{Hammer:2016prh} in a toy model with only one continuum channel
included that, as one increases the coupling
further, something unusual happens (analogous results were reported earlier in Ref.~\cite{vanBeveren:2006ua}):
 While most of the states get stable again and end in
the limit of very large couplings again at masses similar to the original ones,
 typically one state very strongly couples to the continuum channel. The latter kind of state might be viewed as a
 molecular state while the others preserve their $\bar QQ$ nature, however, with changed decay patterns.
The calculations in Ref.~\cite{Hammer:2016prh} were performed for a fixed number of input quark model
states, however, the pattern was observed independent of the number of states involved. In particular:
the state that showed a very strong coupling to the continuum was a mixture of all other states included
regardless of their distance to the actual pole location. Such a collective phenomenon might indicate
the onset of significant $t$--channel exchanges for the binding potential --- after all quark hadron duality
indicates that an infinite sum of $s$--channel poles can be mapped onto an infinite sum of $t$-channel poles.
At the end this provides an understanding how it could be possible that there is a coexistence of quark
model states and molecular states with nearly no cross talk between the two groups.

View III admits the possibility that $\chi_{c1}(4274)$,  $\chi_{c0}(4300)$,  $\chi_{c0}(4500)$
could exist as quark model states even in the presence of lower lying molecular structures.
Analogously, there could be quark model states in the vector channel above 4.4 GeV
even with $Y(4260)$ and $Y(4360)$ being molecular states. At present the $\psi(4415)$ could
be both --- a $D_2\bar D^*$ molecular state as well as a quark model state or a mixture
of both. If indeed there is a mechanism at work as revealed in Ref.~\cite{Hammer:2016prh},
any quark model state that appears above an open $S$--wave threshold should not
decay into this channel. Accordingly, one would expect rather narrow quark model states
even higher up in the spectrum.


\section{\label{Models}Comparison with model calculations}

View I suggests that the $Q\bar Q$ should be screened; the usual Cornell potential with
a Coulomb part proportional to $\alpha_s/r$ and a confinement part linearly rising with
$r$ should be replaced by a Coulomb potential plus a term that could be modelled by $b(1-e^{-\mu r})/\mu$. For
small $r$, the linearly rising potential is reproduced, for large $r$ the potential approaches 
a constant value.
 
Godfrey and Moats~\cite{Godfrey:2015dia} use the classical Cornell potential to calculate the 
bottomonium mass spectrum; in Refs.~\cite{Deng:2016stx,Deng:2016ktl}  the quark-antiquark 
potential is supposed to be flattened by replacing the linear part $br$ of the Cornell potential 
by a screened potential. The comparison of the predicted mass spectrum with 
the experimental masses from the PDG~\cite{Tanabashi:2018oca} in the vector channel 
\bc
\renewcommand{\arraystretch}{1.2}
\begin{tabular}{llllllll}
\cite{Tanabashi:2018oca}                 & 9460  & 10023 & 10355  & 10579  & 10860 & 11020  & MeV \\
\cite{Godfrey:2015dia}                     & 9465  & 10003 & 10345  & 10635  & 10878 & 11102 & MeV \\
\cite{Deng:2016stx,Deng:2016ktl}     & 9460  & 10015 & 10363  & 10597  & 10881 & 10997 & MeV \\
\end{tabular}
\renewcommand{\arraystretch}{1.2}
\ec
shows discrepancies in the order of 27\,MeV for \cite{Godfrey:2015dia}
and 16\,MeV for \cite{Deng:2016stx,Deng:2016ktl}. Better agreement between experiment and quark models 
is achieved when a screened potential is used. 

Table~\ref{comp} shows a comparison of the experimental mass spectrum of charmonium states with
predictions of selected models. A comparison with a large number of other models is shown
in Ref.~\cite{Kher:2018wtv}.  First we discuss the predictions based on the Cornell potential~\cite{Barnes:2005pb,Ebert:2002pp}, which nicely fits to
the first scenario of View II discussed above. The mean deviation of the prediction is 34\,MeV~\cite{Barnes:2005pb}
and 22\,MeV~\cite{Ebert:2002pp}. In the models exploiting the Cornell potential, the $\psi(4040)$ state is identified with
$\psi(3S)$, the $\psi(4415)$ state with $\psi(4S)$. Thus, the classic Cornell potential
reproduces with acceptable precision the spectrum of resonances identified as $Q\bar Q$ mesons in View II.
However, $\psi(4260)$ and $\psi(4360)$ are missing. According to this scenario also the $\chi_{c1}(4160)$ 
should not exist.  

The results using a screened potential~\cite{Li:2009zu,Deng:2016stx} agree with the experimental 
masses with similar accuracy. The mean mass difference is now 26\,MeV~\cite{Li:2009zu} or 29\,MeV~\cite{Deng:2016stx}. 
The $\psi(4S)$ is now expected at 4277\,MeV (mean value of the two models) and identified
with $\psi(4260)$; the $\psi(3D)$ is expected at 4320\,MeV and identified with  $\psi(4360)$. And the
$\psi(4415)$ state, so far identified as $\psi(4S)$, now becomes $\psi(5S)$. The sequence of $\chi_{c1}(nP)$
states can also be mapped with reasonable accuracy onto the experimentally observed states. 
The screened potential provides a natural interpretation
 of the vector states and accommodates the $\chi_{c1}(4160)$. Hence it is somewhat favored.

\begin{table}[htb]
\begin{center}
\caption{Charmonium mass spectrum. The measured masses (MeV) from the PDG~\cite{Tanabashi:2018oca}
are compared to the predictions from  Ref.~\cite{Barnes:2005pb,Ebert:2002pp} using a Cornell
potential and with \cite{Li:2009zu,Deng:2016stx} using a screened potential. The states with masses
in italic are not established. Note that the interpretation of the states $\psi(4260)$,
$\psi(4360)$, and $\psi(4415)$ is different. See text for discussion. }
\label{comp}
\renewcommand{\arraystretch}{1.4}
\begin{tabular}{ccc|ccc|ccc}
\hline\hline
 $n^{2S+1}L_J$ & name &$J^{PC}$ &\cite{Tanabashi:2018oca} &~\cite{Barnes:2005pb} &\cite{Ebert:2002pp}
                                &\cite{Tanabashi:2018oca} & \cite{Li:2009zu}     &\cite{Deng:2016stx}\\
 \hline
$1 ^1S_{0}$    &$\eta_{c}(1S)$   &$0^{-+}$  &2984  &2982 &2981 &2984  &2979 &2984 \\
$2 ^1S_{0}$    &$\eta_{c}(2S)$   &$0^{-+}$  &3638  &3630 &3639 &3638  &3635 &3623\\
$3 ^1S_{0}$    &$\eta_{c}(3S)$   &$0^{-+}$ &\it3940&4043 &3989&\it3940&3991 &4004 \\
$1 ^3S_{1}$    &$J/\psi(1S)$     &$1^{--}$  &3097  &3090 &3096 &3097  &3097 &3097\\
$2 ^3S_{1}$    &$\psi(2S)$       &$1^{--}$  &3686  &3672 &3686 &3686  &3673 &3679\\
$1 ^3D_{1}$    &$\psi_1(1D)$     &$1^{--}$  &3773  &3785 &3783 &3773  &3787 &3792\\
$3 ^3S_{1}$    &$\psi(3S)$       &$1^{--}$  &4039  &4072 &4039 &4039  &4022 &4030\\
$2 ^3D_{1}$    &$\psi_1(2D)$     &$1^{--}$  &4191  &4142 &4150 &4191  &4089 &4095 \\\hline
$4 ^3S_{1}$    &$\psi(4S)$       &$1^{--}$  &4421  &4406 &4427 &4230  &4273 &4281  \\
$3 ^3D_{1}$    &$\psi(3D)$     &$1^{--}$  &  -   & -   &4150 &4368  &4324 &4317 \\
$5 ^3S_{1}$    &$\psi(5S)$       &$1^{--}$  &  -   & -   &  -  &4421  &4463 &4472  \\\hline
$1 ^3P_{2}$    &$\chi_{c2}(1P)$  &$2^{++}$  &3556  &3556 &3555 &3556  &3554 &3553\\
$2 ^3P_{2}$    &$\chi_{c2}(2P)$  &$2^{++}$  &3927  &3972 &3949 &3927  &3937 &3937\\
$1 ^3P_{1}$    &$\chi_{c1}(1P)$  &$1^{++}$  &3511  &3505 &3511 &3511  &3510 &3521\\
$2 ^3P_{1}$    &$\chi_{c1}(2P)$  &$1^{++}$  &3872  &3925 &3906 &3872  &3901 &3914\\
$3 ^3P_{1}$    &$\chi_{c1}(3P)$  &$1^{++}$  &4274  &4271 &4319 &4147  &4178 &4192\\
$1 ^3P_{0}$    &$\chi_{c0}(1P)$  &$0^{++}$  &3415  &3424 &3413 &3415  &3433 &3415 \\
$2 ^3P_{0}$    &$\chi_{c0}(2P)$  &$0^{++}$&\it3862 &3852 &3870&\it3862&3842 &3848\\
$1 ^1P_{1}$    &$h_{c}(1P)$      &$1^{+-}$  &3525  &3516 &3525 &3525  &3519 &3526\\
$1 ^3D_{2}$    &$\psi_{2}(1D)$   &$2^{--}$  &3822  &3800 &3795 &3822  &3798 &3807\\
\hline\hline
\end{tabular}
\end{center}
\end{table}

This result  can be used to argue in favor of View I and in favor of View II:
Screening is a concept inherent in View I. At the first $S$-wave threshold, light quarks are supposed to 
screen the $Q\bar Q$ potential in a particular partial wave. Indeed, the only physics reason for a flattening
of the potential can be the presence of light quarks. Thus states residing in the mass range where the
potential already shows a significant deviation from the Cornell potential already contain light quarks
in their wave function. To more illustrate this point we remind the reader of the Born Oppenheimer
approximation: Here the potential of the heavy nuclei in a molecule is calculated first for the electrons
in the presence of for static nuclei. Once this potential is determined, one can calculate the energy
levels for the nuclei straightforwardly. Obviously the electrons play a crucial role in the molecular binding.
 Something similar is happening here. The ideas of the Born Oppenheimer approximation are transferred
 to doubly heavy systems in Ref.~\cite{Braaten:2014qka} and worked out in more detail 
in Ref.~\cite{Brambilla:2017uyf}. 
On the other hand, from the model parameters one derives that the maximum
total energy in the quark model -- where light-meson loops are neglected -- is  given by
13.193\,GeV for the bottomonium and 4.967\,GeV for the charmonium system. These values are far
above the thresholds for the lowest $S$-wave decays of the vector mesons. In view I, the opening of
thresholds should lower the flattening energy decisively. More definite conclusions on
the performance of screened potentials and the emergence of molecular states can only be drawn
for calculations were the potentials flatten near the $S$--wave thresholds and non--perturbative
meson--meson interactions are included on top.

The spectrum calculated using a screened potential is in very good agreement 
with the second scenario in View II. The screening energies are 
above the thresholds for the lowest $S$-wave decays of all mesons considered here, and the 
screening is felt only as a distortion of the mass of the states: the distortion is small compared to 
the difference between screening energy and the mass of the states involved.
The mass of the highest-mass $\Upsilon$ resonance is shifted by 23\,MeV, the gap to the screening energy
is more than 2\,GeV. The mass predictions for $\psi(4S)$ for the models with screened and unscreened 
potential differ by 140\,MeV, the
 mass gap from $\psi(4415)$ to the screening energy is 550\,MeV. Thus, the screened potential is in View II 
just a modification of the potential pointing at an increasing role of light quarks but with heavy quarks
still playing a decisive role in the formation and dynamics of the heavy quarkonia.

\section{Predictions and data needed}


The interpretation of the states often suffers from the lack of data on decay modes.
Several states are known from one single -- mostly exotic -- channel like $\phi J/\psi$,
$\pi^+\pi^-\psi(2S)$ or $\omega\chi_{c0}$.  Sometimes, one may guess that different sightings
should be combined to a single resonance; in these cases, coupled channel analyses are
needed to substantiate this possibility. In other cases, resonances are seen only in one channel or in
even a multitude of channels like $\psi(4415)$; but in the latter case, most decay modes are {\it seen} only,
and just two finite values for branching ratios are given.

A decisive role for the interpretation is played by the $\psi(4260)$ and $\psi(4360)$ resonances. 
The experimental evidence for these two states is strong and likely,
they are seen in many different channels. However, 
they are incompatible with the first scenario of the View-II $Q\bar Q$ picture. There is the remote
chance that these states might be explained by dynamical effects due to threshold openings and interference effects,
but this is at present a pure speculation. 
An indication that there are light quarks relevant for the formation of these states is that they can be described
within a screened $\bar QQ$ potential. This may be an indication for a molecular character of the states. However, 
the screening energy is rather large and the deviations from a Cornell-type potential may indicate only the
presence of molecular components in the wave function. Also the analysis of Ref.~\cite{Chen:2019mgp}
calls for a significant light quark component in the $Y(4260)$.

Highly important are hence the $e^+e^-$ partial decay widths. 
If $\psi(4260)$, $\psi(4360)$, and $\psi(4415)$ are the $\psi(4S)$, $\psi(3D)$, and $\psi(5S)$ 
states, possibly mixed (second scenario of View II), they should have 
a significant $e^+e^-$ partial decay width.
In this case, the observed decay modes can represent only a small fraction of all decay modes. 
High-statistics data of  $e^+e^-$ scans with many reconstructed final states should be made. 
In View I, little can be said, however, since $e^+e^-$ decays test the short range behavior and 
that can not be quantified  in the molecular picture. 

High-precision profile measurements of the line shapes as suggested for the PANDA experiment
provide sensitive tests of the molecular character of states~\cite{PANDA:2018zjt}. 
However, it needs to be studied which part of the wave functions are probed by those
profiles; a predominantly molecular wave
function at larger distances is possible in View I and in View II.

Presumably, the final answer favoring View I or II needs to come from the pattern of states.
‚In the low-mass region, resonances are found
about 30\,MeV below a threshold. This observation allows us to speculate about the existence of a series of states. 
The comparison with quark-model calculations using a screened potential shows that the density of molecular states could be
larger than the density of quark model states.

So far, no scalar resonance
above $\chi_{c0}(1P)$ is {\it established} and listed in the RPP Summary Table, three candidates
are reported at 3860, 4500 and 4700\,MeV. 


In View I, scalar resonances are in principle possible close to the $D\bar D$, $D_s\bar D_s$,  $D^*\bar D^*$, 
$D_s^*\bar D_s^*$ thresholds at 3930, 3972, 4014, and 4224\,MeV. The two lower-mass scalar 
mesons should not have tensor partners. The $D^*\bar D^*$, $D_s^*\bar D_s^*$ 
thresholds could host both scalar and tensor mesons. In View I, we may thus expect similar masses
for scalar and tensor mesons at about 4014, and 4224\,MeV. Again, the number of possible molecular
states exceeds the number of quark-model states. 

The two presently observed states,
$\chi_{c2}(3930)$ and $\chi_{c0}(3860)$ are difficult to reconcile with these expectations even
though one has to have in mind that $\chi_{c0}(3860)$  is an unconfirmed state.
Quark models predict a mass splitting between tensor and scalar mesons of about  90 -- 100\,MeV 
for $2P$ states and 60 -- 70\,MeV for $3P$ states. At the $2P$ 
level, the experimental mass difference is smaller, $68^{+26+40} _{-32-13}$\,MeV, but not
incompatible.

There are three $\chi_{c1}$ states  above $\chi_{c1}(1P)$. 
The two mesons $\chi_{c1}(4140)$ and  $\chi_{c1}(4274)$ are difficult to accommodate in View I
even though the lower-mass state can possibly be interpreted as $D_s^*\bar D_s^*$ bound state~\cite{Branz:2009yt}. 
Since $\chi_{c1}(3872)$ is interpreted as $D^*\bar D$ molecule, one could (but does not need to) expect a
 $D^*_s\bar D_s$ partner
at 4081\,MeV. In View II, the masses of the $\chi_{c1}$ states are predicted to fall in between the masses 
of their scalar and tensor partners; this expectation holds true for $\chi_{c1}(3872)$ but at present
this expectation cannot be tested for the higher mass states. Nevertheless, the next state is predicted to
have a mass of 4178\,MeV, in fair agreement only with the observed mass.

In View I, two  $h_{c}$ states could be generated dynamically from the $D^*\bar D$ and 
$D^*\bar D^*$ interactions and should show up close to their respective thresholds (although 
at present there are not enough data to fix the potential
for the spin partners of the $\chi_{c1}(3872)$ aka $X(3872)$ and
 to predict if those states should really be bound states or not). 
The quark model predicts the $h_c(2P)$ at about 3900\,MeV
as partner of $\chi_{c2}(2P)$ at 3930\,MeV, $\chi_{c1}(2P)$ at 3872\,MeV, $\chi_{c0}(2P)$ at 3862\,MeV.
 At the $3P$ level, the mass of  $h_c(3P)$ is expected in the 4150 to 4200\,MeV mass region.

The existence of isovector states $Z_c$ and $Z_b$ is in striking conflict with both scenarios of View II. 
The possibility that molecular states (or tetraquarks) exist in partial waves not accessible to $c\bar c$
offers an (unsatisfying) escape.

\section{Summary and Conclusions}
Clearly at present we do not know, if  meson-meson  interactions generate poles or if $\bar QQ$ poles 
drive the non--perturbative part of the hadron-hadron interactions in the quarkonium mass range.
In this article we have examined these two different views in an attempt
to identify experiments or analyses which may be able to decide which view is realised in nature. 

The isovector resonances that decay to states that contain a heavy quark and a heavy
antiquark (either as quarkonium or as a pair of open flavor states) can certainly not be reduced to a simple $Q\bar Q$ structure.  Even though these resonances could certainly 
be generated dynamically from their decay particles, the possibility so far persists that they could be produced via
perturbative rescatterings in the final state and interpretations of these states as kinematical enhancements
cannot yet be ruled out. Fortunately this issue can be resolved once better data are available.

Precision experiments scanning the resonance region in $e^+e^-$ annihilation
should reveal   the number of $\psi$ state and the electronic partial decay widths of the vector mesons; this information can help
to decide on the nature of vector states above 4\,GeV. Alternative interpretations of  $\psi(4260)$ and $\psi(4360)$ 
should be scrutinized. A sensitive search for scalar and tensor mesons may 
shed light onto the pattern of states. The molecular picture of the high-mass states links these
expected states to the thresholds of opening channels which are partly identical for scalar and tensor
mesons. Quark models predict a hierarchy: tensor mesons are higher in mass than their scalar partners.

The molecular view allows for two $h_c$ states in the 3800 to 4100 MeV mass range. The quark model
suggests that the masses should be 3900\,MeV and between 4150 and 4200\,MeV. 

Finally, it may turn out that the concepts we learned in heavy-meson spectroscopy are not easily extendable
to the physics of light quarks. 
Since for the heavy quark sector QCD is probed in parts in its perturbative regime with $\alpha_s$ being small, 
while its  non-perturbative regime is probed in the light quark sector, there is  no guarantee that the
transition from the one to the other is smooth. But there is still the hope that
both fields can learn from each other and that a common understanding will finally emerge.

{\it\small
This work was triggered by a lecture series at PNPI (Kurchatov Institute),
Gatchina, by E. K. We acknowledge support by the Deutsche Forschungsgemeinschaft (SFB/TR110).
This work is partially supported
by the National Natural Science Foundation of China (NSFC) and Deutsche Forschungsgemeinschaft (DFG) through 
funds provided to the Sino--German Collaborative Research Center ``Symmetries and the
Emergence of Structure in QCD'' (NSFC Grant No.~11621131001,
DFG Grant No.~TRR110).}


\begin{thebibliography}{99}
\bibitem{footnote2}We here use the spectroscopic name
for the states whenever the PDG uses it.
\end{thebibliography}
\end{document}